\documentclass[twocolumn,aps,amsmath,amssymb]{revtex4-1}

\usepackage{natbib}
\usepackage{hyperref}
\usepackage{bm}
\usepackage{graphicx}
\usepackage{dcolumn}
\usepackage{bm}
\usepackage{xcolor}
\newcommand{\gtp}{g_{t\phi}}
\newcommand{\gtt}{g_{tt}}
\newcommand{\gpp}{g_{\phi \phi}}
\newcommand{\grr}{g_{rr}}
 \newcommand{\gthth}{g_{\theta \theta}}

\renewcommand{\tt}{\tilde{T}}

\newcommand{\rin}{r_{\text{in}}}
\newcommand{\rout}{r_{\text{out}}}

\newcommand{\gab}{g^{\alpha \beta}}
\newcommand{\goab}{g_{\alpha \beta}}
\newcommand{\Tab}{T^{\alpha \beta}}
\newcommand{\Tabmat}{T^{\alpha \beta}_{\text{MAT}}}
\newcommand{\Tabem}{T^{\alpha \beta}_{\text{EM}}}
\newcommand{\Fab}{F^{\alpha \beta}}
\newcommand{\Fabext}{F^{\alpha \beta}_{\text{EXT}}}
\newcommand{\Fabint}{F^{\alpha \beta}_{\text{INT}}}

\newcommand{\At}{A_{t}}
\newcommand{\Ap}{A_{\phi}}
\newcommand{\drp}{\partial_{r}p}
\newcommand{\dtp}{\partial_{\theta}p}
\newcommand{\drh}{\partial_{r}h}
\newcommand{\dth}{\partial_{\theta}h}
\newcommand{\dr}{\partial_{r}}
\newcommand{\dt}{\partial_{\theta}}
\newcommand{\drr}{\partial^2_{rr}}
\newcommand{\dtt}{\partial^2_{\theta\theta}}

\usepackage{float}
\def\aap{{A\&A}} 
\def\mnras{{MNRAS}} 
\def\jcap{{J. Cosmology Astropart. Phys.}}
\def\apjs{{ApJS}} 
\def\aapr{A\&A~Rev.}          
\def\araa{ARA\&A}             
\newcommand{\BibitemShut}[1]{}

\begin{document}


\title{Equilibrium configurations of charged fluid around Kerr black hole}


\author{Audrey Trova}
	\email{audrey.trova@zarm.uni-bremen.de}
\author{Kris Schroven}
\author{Eva Hackmann}
\affiliation{University of Bremen, Center of Applied Space Technology and Microgravity (ZARM), 28359 Bremen, Germany}

\author{Vladim\'{i}r Karas}
\affiliation{Astronomical Institute, Academy of Sciences,
Bo\v{c}n\'{i} II, CZ-141\,31\,Prague, Czech Republic}

\author{Ji\v{r}\'{i} Kov\'{a}\v{r}}
\author{Petr Slan\'{y}}
\affiliation{Institute of Physics, Faculty of Philosophy and Science,
Silesian University in Opava\\ Bezru\v{c}ovo n\'{a}m. 13, CZ-746\,01 Opava, Czech Republic}%


\date{\today}

\begin{abstract}
Equilibrium configurations of electrically charged perfect fluid surrounding a central rotating black hole endowed with a test electric charge and embedded in a large-scale asymptotically uniform magnetic field are presented. Following our previous studies considering the central black hole non-rotating, we show that in the rotating case, conditions for the configurations existence change according to the spin of the black hole. We focus our attention on the charged fluid in rigid rotation which can form toroidal configurations centered in the equatorial plane or the ones hovering above the black hole, along the symmetry axis. We conclude that a non-zero value of spin changes the existence conditions and the morphology of the solutions significantly. In the case of fast rotation, the morphology of the structures is close to an oblate shape. 
\end{abstract}


\maketitle


\section{Introduction}
In active galactic nuclei, accretion of matter from inner accretion discs leads to intense emission of X-rays \citep{Pado17}. The emerging energetic radiation then irradiates the cooler outer thick discs, where the temperature drops below the critical value of the dust sublimation. Dust grains acquire electric charge due to photoelectric effects and the complex plasma environment \citep{Mendis94,Horanyi96}. At the same time, the continued accretion events causes the central black hole to spin up. Therefore, the effects of fluid charge and rotation of the central black hole need to be taken into account together.

The thick accretion discs, especially those with negligible loss of mass, can be well modeled by the toroidal equilibrium configurations (tori) of fluids studied in proper gravito-electromagnetic backgrounds. Starting with the basic Schwarzschild and Kerr ones \cite{AbrJaSi78,KoJaAbra78}, many studies of neutral perfect fluid tori were realized in more complicated backgrounds \cite{FrKinRain02,FontDaigne02,StuSlaHle00,RezZanFon03,SlaStuc05,KucSlStu11,Pugliese13}, also with the presence of the background magnetic fields \cite{Komissarov06,KovarTr14,KoSlaCreStuKaTro16}. All of these studies were performed in the general relativistic framework, but many others also exist in the Newtonian description \cite{CreKoSlaStuKa13,slany13}. Thus, dozens of scenarios of rotating fluids have been described by taking many factors into account; some of them including the self-magnetic fields generated by the moving charged fluid as well. However, a consideration of the self-gravitational field of the torus is more or less rare \cite{OstMa68,Eriguchi05,OtaTakEri09,Trova16}. On the other hand, this self-field can be mostly safely neglected for low-mass toroidal structures. 

The purpose of this paper is an investigation of charged perfect fluids encircling rotating black hole endowed with an electric charge and embedded in a large-scale asymptotically uniform magnetic field. This background represents a `rotational' generalization of the static one considered in our previous paper \cite{KovarTr14}. Here, the strong gravity near the considered rotating black hole is described by the Kerr spacetime, which is combined with the test external electromagnetic field, describing the test charge of the black hole and the asymptotically uniform magnetic field. 
Let us remind the reader that the test electric charge itself is likely to converge to a very small value \citep{Wald74,Azreg16}. Despite this fact, electric charge distribution emerges within the surrounding environment; the charges can separate and play a significant role in the interaction with the global magnetic field.

Even if the considered gravitational and electromagnetic background generally plays an important role in the structure of accretion flows (gravitational fragmentation \cite{ParBla03,SalSiAe16}, thermal stabilization of the flow \cite{Sadowski16}), our interest here is the impact of the rotation of the black hole on the topology of the fluid toroidal structures, comparing them with the ones in the non-rotating background \cite{KovarTr14}.
According to the forces acting on the system, fluids can exhibit various types of behavior. Here, we focus our study on two different types of configurations: the regular ones, centered in the equatorial plane (toroidal configurations), and the unique ones, located along the symmetry axis (polar cloud configurations). 

In active galactic nuclei, strong irradiation of a dusty torus by X-rays originates from the central source and it leads to charging of grains due to photoelectric effect \citep{Kovar11,Trova16}. It can be shown that the mechanism of dust charging helps to levitate grains above the equatorial plane, thus counter-acting the collapsing role of self-gravity in magnetized discs and tori. The interplay between the gravitational attraction and the repulsion/attraction of the electromagnetic effect is important to characterize the motion of matter. It has been shown \cite{KoKoKaStu10} that halo orbits (off-equatorial circular orbits) of electrically charged particles exist near compact objects. This leads us to study the equatorial tori which can represent the accretion disc, and the polar clouds that can scatter and polarize light on the axis \citep{Antonucci93,MarMa96}
While the central black hole dominates the gravitational field and remains electrically (almost) neutral, the surrounding material has a non-negligible self-gravitational effect on the torus structure and, moreover, by charging mechanisms it acquires non-zero electric charge density. While these influences need to be taken into account together in order to achieve a self-consistent picture, in the present paper we concentrate on the latter one (charge distribution), which appears to be of primary role on the structure of equilibrium configurations.

These structures can be constructed within the model following from the conservation laws and Maxwell equations, where, along with the basic assumptions of the axial symmetry and stationarity of the background, other assumptions are imposed. 
Especially, we consider the fluid in pure azimuthal motion and with adherent charges, thus moving convectively together with the fluid only (zero conductivity of the fluid -- the opposite limit of the ideal magnetohydrodynamics requirement);  the structures are modeled by the fluid with a non-vanishing electric charge density, such as an ionized plasma \cite{WarNg99,InouInu08,PandWrd08}. Moreover, radiation and viscosity of the moving fluid are neglected, as it is characteristic for the perfect fluids.
In order to calculate the desired pressure profiles, we also need to specify the angular velocity or the angular momentum profiles of the circling fluid. Here, we chose the profile of constant angular velocity (rigid rotation);  even if the real discs have more complicated velocity profiles, the rigid rotation has the benefit that the problem can be treated analytically. To close the system of equations, we specify the fluid obeying a polytropic equation of state. Finally, we neglect the self-gravitational and self-electromagnetic fields produced by the circling charged fluid.

The paper is organized as follows. In Sec. \ref{sec:section1}, we describe the background of the system. The Kerr spacetime metric is introduced and the physical characteristics of the fluid, such as density, angular momentum and velocity are prescribed. In this section, we also present the basic equations of the systems. Sec. \ref{sec:IntCondExCond} is dedicated to the integrability conditions of the pressure equations and to the main assumptions. The general conditions of equilibrium existence are given in Sec. \ref{sec:section4} and applied to the polar cloud configuration and to the equatorial toroidal configurations. Sec. \ref{sec:section5} is devoted to the construction of the polar cloud structures and equatorial tori. Various solutions are presented as a function of the spin of the black hole. We focus our attention on two particular cases: a fast rotating and a slowly rotating black hole. We conclude our work with a study of two limiting cases: the zero electric charge of the black hole and the zero strength of the external magnetic field.   

All along the paper, for quantities denoted by a bar, $\overline{x}$, we use the geometrical system of units. They become dimensionless, $x$, when they are scaled by the mass of the black hole. Finally, for a direct interpretation, we express them in physical units (SI) and denote them as $\tilde{x}$.

\section{\label{sec:section1}Framework}
\subsection{Gravitational field: Kerr metric}
The considered fluid tori have a negligible mass and they do not contribute to the gravitational field, which is generated by the Kerr black hole. The considered Kerr spacetime is described, in the Boyer-Linquist spheroidal coordinates $(t,r,\theta,\phi)$, by the axially symmetric metric tensor $\goab$, written in the dimensionless system of units in the form 
\begin{align}
\nonumber
&\gtt=-\left(1-\frac{2r}{\Sigma}\right)\,,\\ 
\nonumber
&\gtp=-\frac{2r}{\Sigma}a\sin^2 \theta\,,\\ 
&\grr=\frac{r^2-2r+a^2}{\Sigma}\,,\\ 
\nonumber
&\gthth=\frac{1}{\Sigma}\,,\\ 
\nonumber
&\gpp=\left(r^2+a^2+\frac{2ra^2}{\Sigma}\sin^2\theta\right)\sin^2\theta\,,
\nonumber
\end{align}
where $\Sigma=r^2+a^2\cos^2 \theta$, $a$ is the dimensionless Kerr spin parameter ($0\leq a \leq 1$).

\subsection{External electromagnetic field}
Along with the gravitational field of the Kerr black hole, the studied toroidal structures are embedded into external asymptotically uniform magnetic field. In terms of the vector potential, this scenario can be described by Wald's test-field solution of the Maxwell equations \citep{Wald74,Stuchlík2016} 
\begin{subequations} \label{eq:VectorPot}
\begin{gather}
\At = \frac{B}{2}\Big{(}\gtp+2a\gtt-e\left(\gtt+1\right)\Big{)}\,, \label{eq:VectorPot1} \\
\Ap = \frac{B}{2}\left(\gpp+2a\gtp-e\gtp\right) \label{eq:VectorPot2}\,,
\end{gather}
\end{subequations}
where $e=Q/B$ is the electromagnetic parameter. The quantities $B$ and $Q$ represent, respectively, the strength of the external uniform magnetic field and the charge of the black hole; as the parameters of the test field, they do not influence the spacetime geometry.
The related Faraday electromagnetic tensor $\Fabext$ can be then determined from the general expression 
\begin{equation}
\Fabext = g^{\alpha\mu} g^{\beta \nu}(\nabla_{\mu}A_{\nu}-\nabla_{\nu}A_{\mu})\,.
\end{equation}

\subsection{Fluid: density, angular momentum and velocity \label{sec:CondPPos}}
\begin{figure}
\begin{tabular}{cc}
\includegraphics[width=.45\hsize]{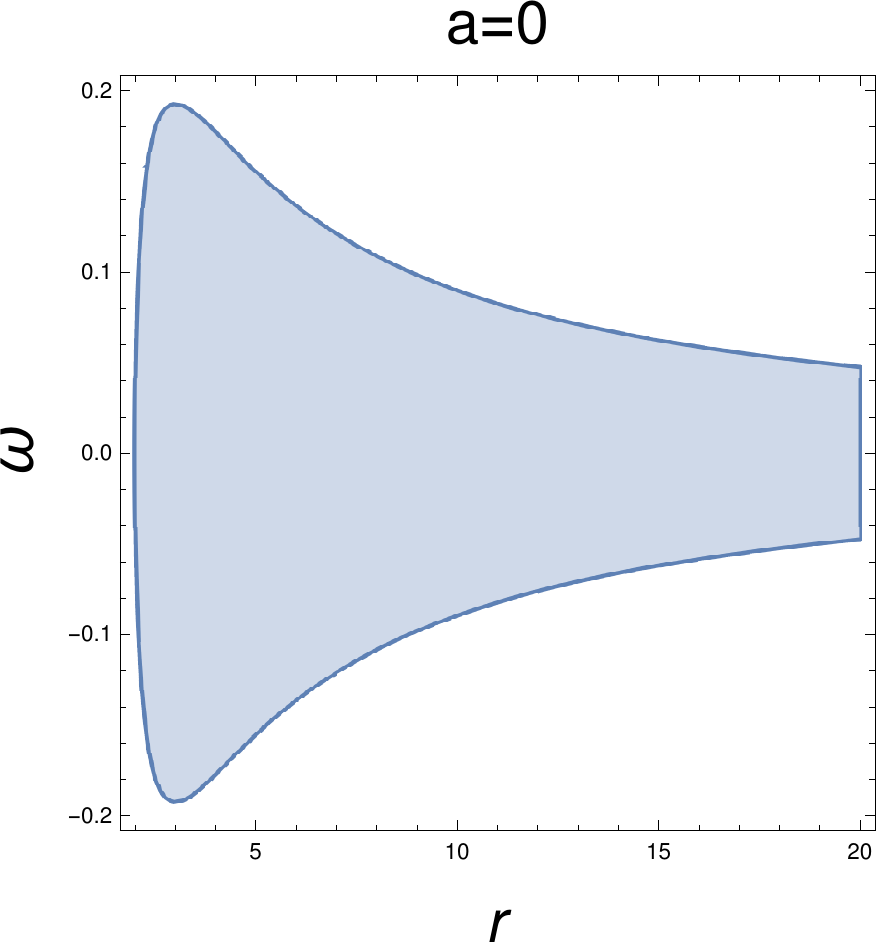} &  \includegraphics[width=.45\hsize]{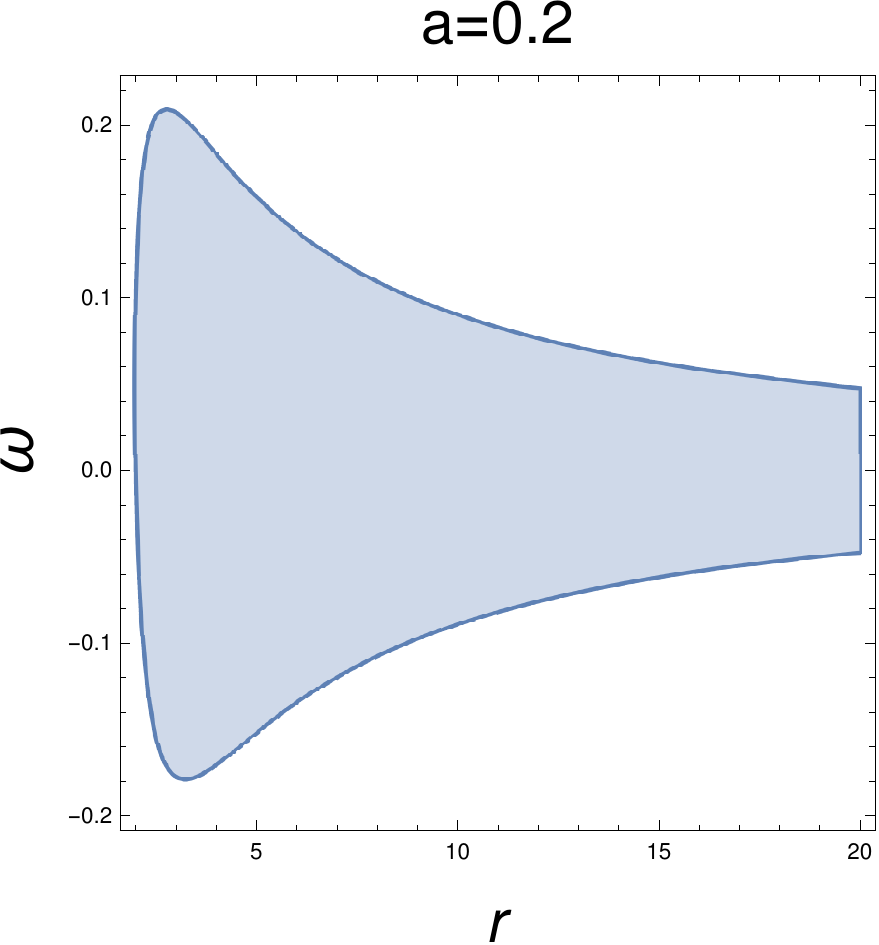} \\
\includegraphics[width=.45\hsize]{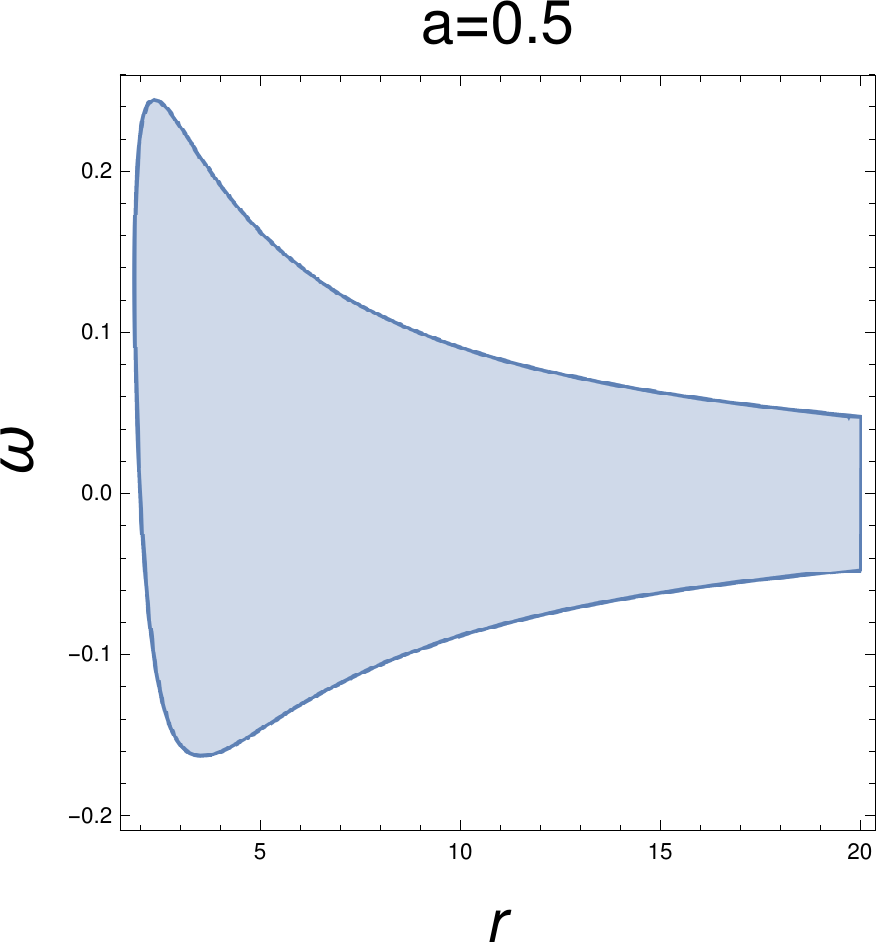} &  \includegraphics[width=.45\hsize]{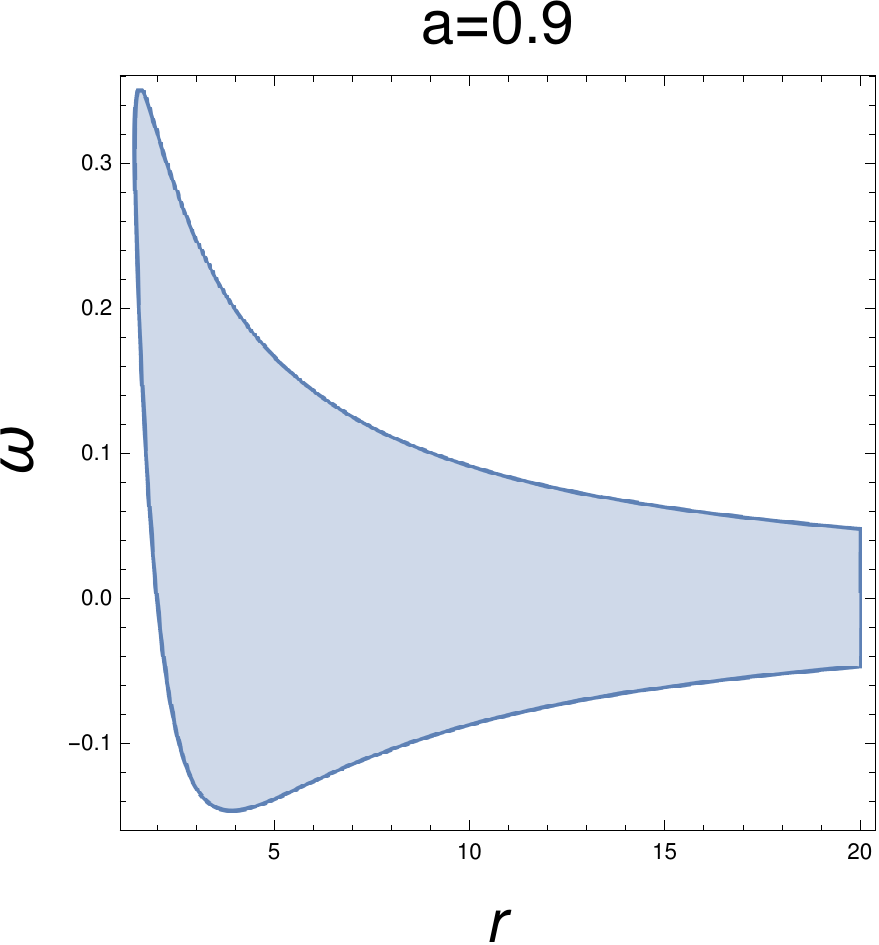} \\
\end{tabular}
	\caption{\label{f1} The shaded region represents the area where the condition ${\cal{P}}>0$ is satisfied in the equatorial plane ($\theta=\pi/2$). The condition is plotted for four values of the spin $a$. An effect of frame dragging can be seen in the shown plots. The mirror symmetry along $\omega=0$ is broken for non-vanishing spin values, and close to the black hole horizon higher absolute values for $\omega$ are allowed in the co-rotating case. This effect becomes larger the bigger the value of $a$ is.}
\end{figure}

Within our model, the perfect fluid is characterized by its motion in the $\phi$-direction only, described by the four-velocity $U^\alpha=(U^t,0,0,U^\phi)$, where
\begin{equation}
(U_t)^2=\frac{(\omega \gtp +\gtt)^2}{{\cal{P}}}\,,
\end{equation}
with
\begin{equation}
{\cal{P}}=-\left(\omega^2 \gpp+2\omega \gtp+ \gtt\right)\,.
\end{equation}
We introduce the specific angular momentum, $l\equiv l(r,\theta)$, and the angular velocity, $\omega \equiv \omega(r,\theta)$, through the usual definitions
\begin{equation}
\omega=\frac{U^{\phi}}{U^t} \quad \text{and} \quad l=-\frac{U_{\phi}}{U_t}\,,
\end{equation}
with $U^t=\sqrt{{\cal{P}}}^{-1}$.
Both quantities are linked to each other by the relations 
\begin{equation}
\omega=-\frac{l\gtt + \gtp}{l\gtp +\gpp},\quad l=-\frac{\omega \gpp + \gtp}{\omega \gtp + \gtt}\,.
\end{equation}
The quantity $(U^t)^2$ provides us with the basic necessary existence conditions for our circling tori. Since $(U^t)^2>0$ then ${\cal{P}}>0$ and it gives us:
\begin{itemize}
\item restriction on the radial coordinate $r>(1+\sqrt{1-a^2}) \equiv r_{H}$, where $r_{\text{H}}$ is the radius of the outer event horizon,
\item restriction on the angular velocity $\omega$ (thus, also on the size of torus), for $\theta=\pi/2$ shown in Fig.~\ref{f1}. 
\end{itemize}
Note that in the case of $\theta=0$, the condition ${\cal{P}}>0$ reduces to $-g_{tt}>0$ which on the symmetry axis coincides with $r > r_{H}$. Therefore, in this case no additional restriction on $\omega$ arises.

\subsection{Pressure equation}
Pressure equations describing a rotating perfect fluid equilibrium configuration can be derived from the following conservation laws and Maxwell equations

\begin{subequations} \label{eq:Maxwell}
  \begin{gather}
\nabla_{\beta} \Tab =0 \label{eq:Continuity1}\,, \\
\nabla_{\beta} \Fab =4 \pi J^{\alpha} \label{eq:Maxwell2}\,,
  \end{gather}
\end{subequations}
where $J^{\alpha}$ is the four-current density which is linked to the four-velocity $U^{\alpha}$, charge density $q$, and electrical conductivity $\sigma$ through the Ohm law,
\begin{equation}
J^{\alpha}=qU^{\alpha}+\sigma \Fab U_{\beta}\,.
\end{equation}
Our assumption of zero conductivity ($\sigma=0$) leads to $J^{\alpha}=qU^{\alpha}$. Moreover, in our scenario of the fluid circulation in the azimuthal direction only, the four-current has the only non-vanishing components
\begin{equation}
J^{\phi}=qU^{\phi} \quad \text{and} \quad J^t=qU^t\,. \label{eq:fourcurrent}
\end{equation}

The total electromagnetic tensor $\Fab$  can be split into two terms
\begin{equation}
\Fab = \Fabext +\Fabint\,,
\end{equation}
where $\Fabint$ describes the electromagnetic self-field produced by the charged fluid. According to our assumption of a test-charged fluid, this self-electromagnetic field (as well as the self-gravitational field) generated by the fluid itself are neglected in comparison with the external one, i.e.$\Fabext \gg \Fabint$, thus, we can write $\Fab \approx \Fabext$. 

The stress-energy tensor $\Tab$ can be also split into two parts 
\begin{equation}
\Tab = \Tabmat +\Tabem\,,
\end{equation}
where $\Tabmat$ is the matter part and given by
\begin{equation}
\Tabmat = (\epsilon + p)U^{\alpha}U^{\beta}+p\gab\,,
\end{equation}
where $p$ is the pressure of the fluid and $\epsilon$ the energy density. The electromagnetic part $\Tabem$ can be expressed as
\begin{equation}
\Tabem = \frac{1}{4\pi}\left(F^{\alpha}_{\;\;\gamma}F^{\beta \gamma}-\frac{1}{4}F_{\gamma \delta}F^{\gamma \delta}\gab\right)\,.
\end{equation}

From the continuity law \eqref{eq:Continuity1}, Maxwell equations \eqref{eq:Maxwell2} and the stress-energy tensor decomposition, we get the following equation \citep{1973MTW}
\begin{equation}
\label{eq:FinalEqCompact}
\nabla_{\beta} \Tabmat = \Fabext J_{\beta}\,,
\end{equation}
where $J_{\beta}$ is the four-current density \eqref{eq:fourcurrent} produced by the charged fluid. Equation \eqref{eq:FinalEqCompact} can be decomposed and this provides us with two differential pressure equations:
\begin{widetext}
\begin{align}
\label{eq:pressureEq}
\nonumber
&\drp = -(p+\epsilon)\left(\dr \ln \vert U_t\vert - \frac{\omega \dr l}{1-\omega l}\right) + q\left(U^t\dr \At+U^{\phi}\dr \Ap\right)\,, \\ 
&\dtp = -(p+\epsilon)\left(\dt \ln \vert U_t\vert - \frac{\omega \dt l}{1-\omega l}\right) + q\left(U^t\dt \At+U^{\phi}\dt \Ap\right)\,.
\end{align}
\end{widetext}
\subsection{Transformation of the pressure equations}
To simplify the pressure equations \eqref{eq:pressureEq} and to avoid a numerical integration, we transform the two non-linear pressure equations to linear ones through the same process as in \citep{KovarTr14}. For this purpose,
we assume that the fluid is described by a polytropic equation of state
\begin{equation}
\label{eq:pressure}
p=\kappa \rho^{\Gamma}\,,
\end{equation}
with $\Gamma$ and $\kappa$ being the polytropic exponent and coefficient, respectively, and $\rho$ is the rest-mass density of the fluid; moreover, we assume that for low temperatures ($p \ll\rho $), the fluid energy density can be approximated as
\begin{equation}
\label{eq:eps}
\epsilon \simeq \rho\,,
\end{equation}
consequently, $\epsilon +p \sim \rho$.
Then, we set
\begin{equation}
\label{eq:K}
K=\frac{q}{\epsilon + p} {\bm{ \simeq \frac{q}{\rho}} }\,,
\end{equation}
and introduce the function $h$ satisfying the relations
\begin{align}
\label{eq:hLinkp}
&\drh = \frac{\Gamma-1}{\Gamma}\frac{\drp}{p+\epsilon} {\bm{ \simeq \frac{\Gamma-1}{\Gamma}\frac{\drp}{\rho}} }\,,\\ 
&\dth = \frac{\Gamma-1}{\Gamma}\frac{\dtp}{p+\epsilon}{\bm{ \simeq \frac{\Gamma-1}{\Gamma}\frac{\dtp}{\rho}} }\,.\nonumber
\end{align}
Here, $K$ is the so-called correction function and it determines the charge density distribution. A proper form of this function ensures the integrability of the pressure equations \eqref{eq:pressureEq}, which can now be written as
\begin{widetext}
\begin{align}
\label{eq:partialh}
\nonumber
&\drh = -\frac{\Gamma-1}{\Gamma}\left[\dr \ln \vert U_t\vert - \frac{\omega \dr l}{1-\omega l}- K\left(U^t\dr \At+U^{\phi}\dr \Ap\right) \right] , \\ 
&\dth = -\frac{\Gamma-1}{\Gamma}\left[\dt \ln \vert U_t\vert - \frac{\omega \dt l}{1-\omega l}- K\left(U^t\dt \At+U^{\phi}\dt \Ap\right) \right] .
\end{align}
\end{widetext}
\section{Integrability conditions \& physical characteristics \label{sec:IntCondExCond}}
\subsection{Correction function \label{sec:IntCon}}
The next step is to integrate the Eqs. \eqref{eq:partialh} and obtain the $h$-function. This set of equations is not in general integrable and depends on parameters, such as the spin parameter $a$, magnetic field strength $B$, electromagnetic parameter $e$, correction function $K$ (linked to the charge density profile $q$) and the specific angular momentum $l$ or angular velocity $\omega$. The function $K$ can be constrained by the integrability conditions of the two partial differential equations \eqref{eq:partialh}. This is given as
\begin{equation}
\label{eq:IntCondGen}
\dt \drh = \dr \dth\,.
\end{equation}
Equation \eqref{eq:IntCondGen} is automatically satisfied for $q=0$, i.e. for a rotating uncharged perfect fluid \citep{KoJaAbra78,AbrJaSi78}. For $q\neq 0$, the integrability condition  is not automatically fulfilled for a given $q$ and $\omega$ or $l$. Thus, in a charged case, the following equation has to be satisfied:\\
\footnotesize{
\begin{equation}
\dt \left[ K(U^t\dr \At+U^{\phi}\dr \Ap)\right]  = \dr \left[ K(U^t\dt \At+U^{\phi}\dt \Ap)\right].
\end{equation}}
\normalsize
To find an analytical solution for $K$, $\omega$ can be set to a constant, meaning that the fluid rotates rigidly. This assumption leads to the equation
\begin{equation}
\label{eq:intcond}
\dt (KU^t) \dr(\At+\omega \Ap)=\dr (KU^t) \dt(\At+\omega\Ap)\,,
\end{equation}
which is satisfied for
\begin{equation}
KU^t=f(S) \quad \text{with} \quad S=\At+\omega \Ap\,.
\end{equation}
Here, $f(S)$ is an arbitrary function that determines the charge density distribution of the rigidly rotating tori (see also \cite{KoSlaStuKaTro17} for an alternative solution of equation \eqref{eq:intcond}). The two partial differential equations \eqref{eq:partialh} can be rewritten as
\begin{align}
\label{eq:drhFin}
\nonumber
&\drh = -\frac{\Gamma-1}{\Gamma}\left[\dr \ln \vert U_t\vert - \frac{\omega \dr l}{1-\omega l}- f(S)\dr S \right] , \\ 
&\dth = -\frac{\Gamma-1}{\Gamma}\left[\dt \ln \vert U_t\vert - \frac{\omega \dt l}{1-\omega l}- f(S)\dt S \right] ,
\end{align}
and integrated with the $h$-function expressed as
\small{
\begin{equation}
\label{eq:hFunction}
h=\frac{\Gamma-1}{\Gamma}\left(-\ln \vert U_t \vert-\ln \vert 1-\omega l \vert+\int{f(S) dS}+h_0\right) . 
\end{equation} 
}
\normalsize
Here, $h_0$ is a constant of integration, which determines the size of the torus (i.e its boundary).

\subsection{Physical characteristics}
From Eqs. \eqref{eq:pressure}--\eqref{eq:hLinkp}  and \eqref{eq:hFunction}, we can determine the rest-mass density and the specific charge distributions
\begin{align}
\label{eq:density}
&\rho = \bm{ \left(\frac{h}{\kappa}\right)^{\frac{1}{\Gamma-1}}} ,\\
\label{eq:specificCharged}
&q_s = \bm{ \frac{q}{\rho} = K.}
\end{align}
The boundary of the torus is then determined by the zero-surface of $p$, $\rho$ or $h$.

\section{Existence conditions \label{sec:section4}}
\subsection{General conditions}
\label{sec:section41}
We investigate the existence of rotating charged configurations which center at the coordinates ($r_c$,$\theta_c$), corresponding to the maxima of the $h$-function. The necessary conditions for a maximum read
\begin{subequations} \label{eq:ExtremumCond}
  \begin{gather}
\dr h\vert_{r=r_c,\theta=\theta_c} = 0\,, \label{eq:ExtremumCond1} \\
\dt h\vert_{r=r_c,\theta=\theta_c} = 0\,, \label{eq:ExtremumCond2}
  \end{gather}
\end{subequations}
while the sufficient ones require in addition
\begin{subequations} \label{eq:MaxCond}
  \begin{gather}
\drr h\vert_{r=r_c,\theta=\theta_c} < 0\,, \label{eq:MaxCond1} \\
\text{det} {\cal{H}}\vert_{r=r_c,\theta=\theta_c} > 0\,, \label{eq:MaxCond3}
  \end{gather}
\end{subequations}
where ${\cal{H}}$ is the Hessian matrix, 
\begin{eqnarray}                                                              
\label{Hessian}
\mathcal{H} =
\left( 
\begin{array}[c]{cc}
\partial^2_{rr} h & \partial^2_{r\theta} h\\
\partial^2_{\theta r} h & \partial^2_{\theta\theta} h
\end{array}
\right).
\end{eqnarray}
Particularly, in this work, we are interested in two configurations:
\begin{itemize}
\item structures centered on the polar axis named ``polar clouds" ($\theta_c=0$),
\item equatorial tori centered in the equatorial plane ($\theta_c=\pi/2$).
\end{itemize}
In both cases, the cross derivatives vanish, thus the condition \eqref{eq:MaxCond3} reduces to
\begin{equation}
\dtt h\vert_{r=r_c,\theta=\theta_c} < 0\,.\label{eq:MaxCond2}
\end{equation} 
The construction of equilibrium configurations depends on the following free parameters: $r_c$, $a$, $B$, $e$, $\omega$ and the function $f(S)$. As for the $f(S)$ function, we use the same forms as in \cite{KovarTr14}, $f(S)=kS^n$, where $k$ is a constant and $n=1$ or $n=-2$, mainly because we know solutions can be found for these choices. Another advantage is that we can establish a comparison between the case of a non-rotating black hole and our case with rotation.

\subsection{Polar cloud case \label{sec:ConditionPolarCloud}}

Our first interest is to focus on structures centered on the polar axis. Before we proceed to analyse the necessary and sufficient conditions for the existence, we have a look on the symmetry of these structures. On the axis $\theta=0,\pi$ the only nonvanishing component of the electromagnetic field tensor is
\begin{align}
 F_{rt}\big|_{\theta=0,\pi} = \frac{(Q-2aB)(r^2-a^2)}{(r^2+a^2)^2}\,.
\end{align}
This implies that on the polar axis an observer does not see a magnetic field but only a radial electric field \citep{Wald74}. Therefore, it is clear that the polar cloud structures we are looking for are symmetric with respect to the equatorial plane.

\begin{figure}
\centering
  \includegraphics[width=1.\hsize]{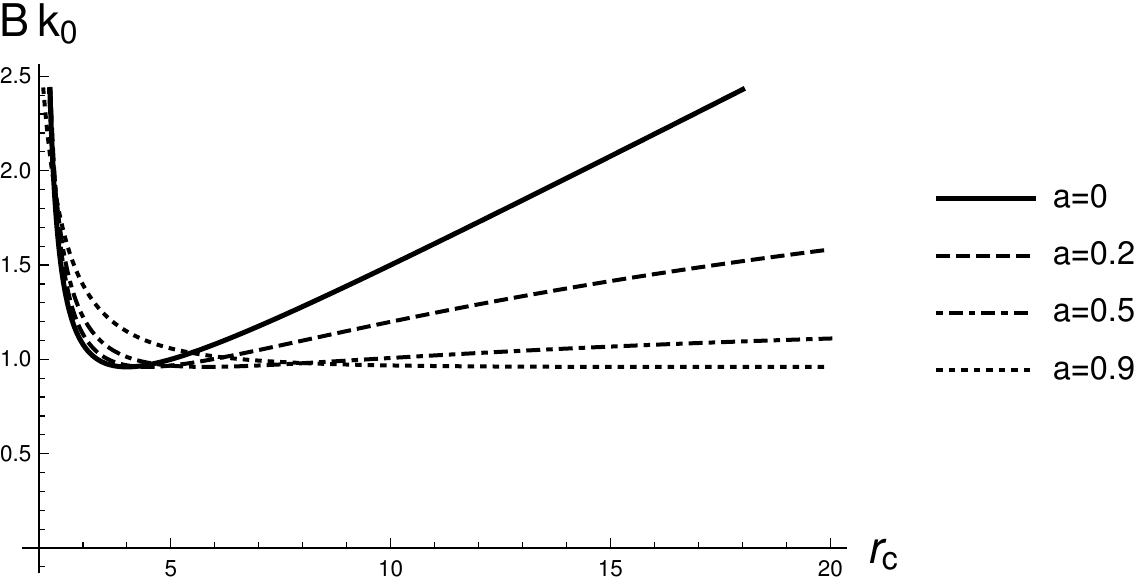}
	\caption{\label{f2} Behavior of $Bk_0(r_c,a,e)$ in function of $r_c$ for four values of the spin $a=0,0.2,0.5,0.9$ and for $e=4.17$.}
\end{figure}

\begin{figure}
\centering
  \includegraphics[width=0.7\hsize]{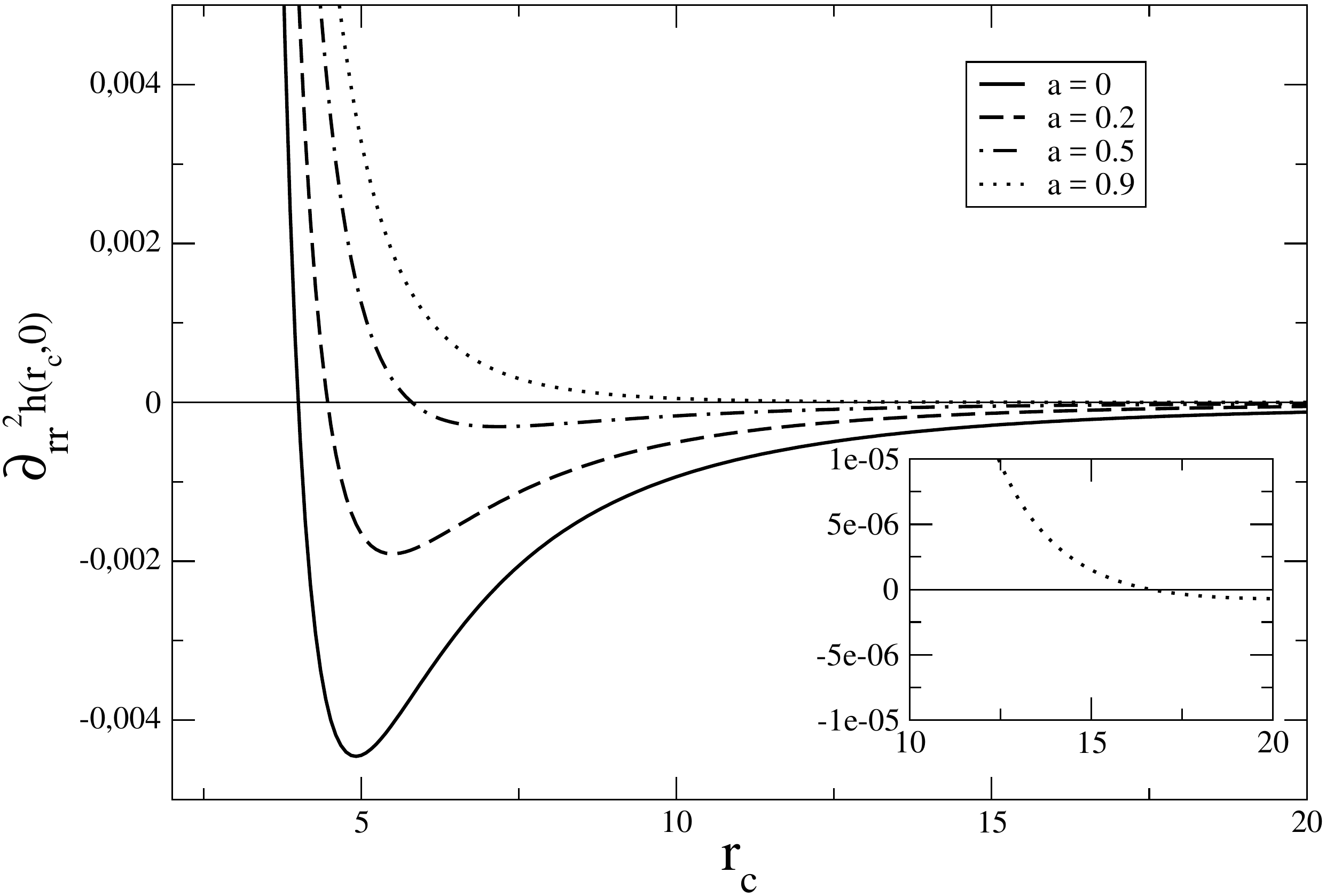}
	\caption{\label{f3} Behavior of the function $\drr h(r_c,0,a,e)$ for four values of the spin $a=0,0.2,0.5,0.9$, for $e=4.17$ and $f(S)$ given by relation \eqref{eq:fpol}. The inserted graph represents a zoom of the case $a=0.9$.}
\end{figure}

\begin{figure}
\begin{tabular}{cc}
\includegraphics[width=.45\hsize]{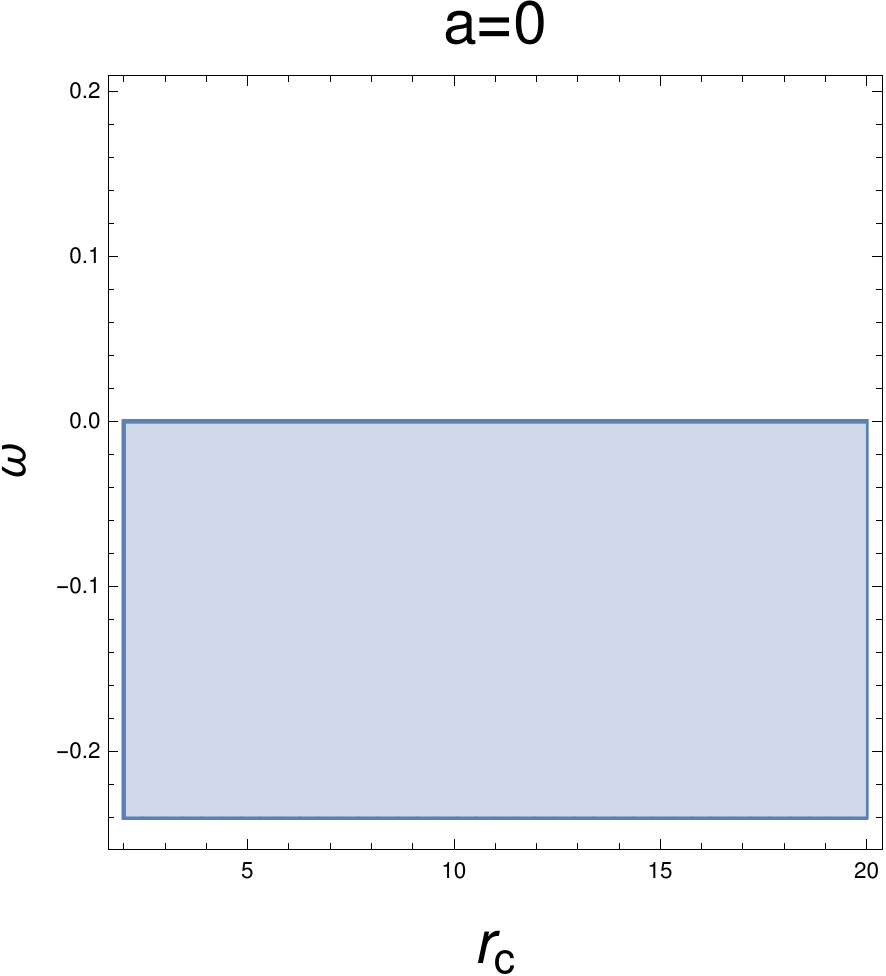} &  \includegraphics[width=.45\hsize]{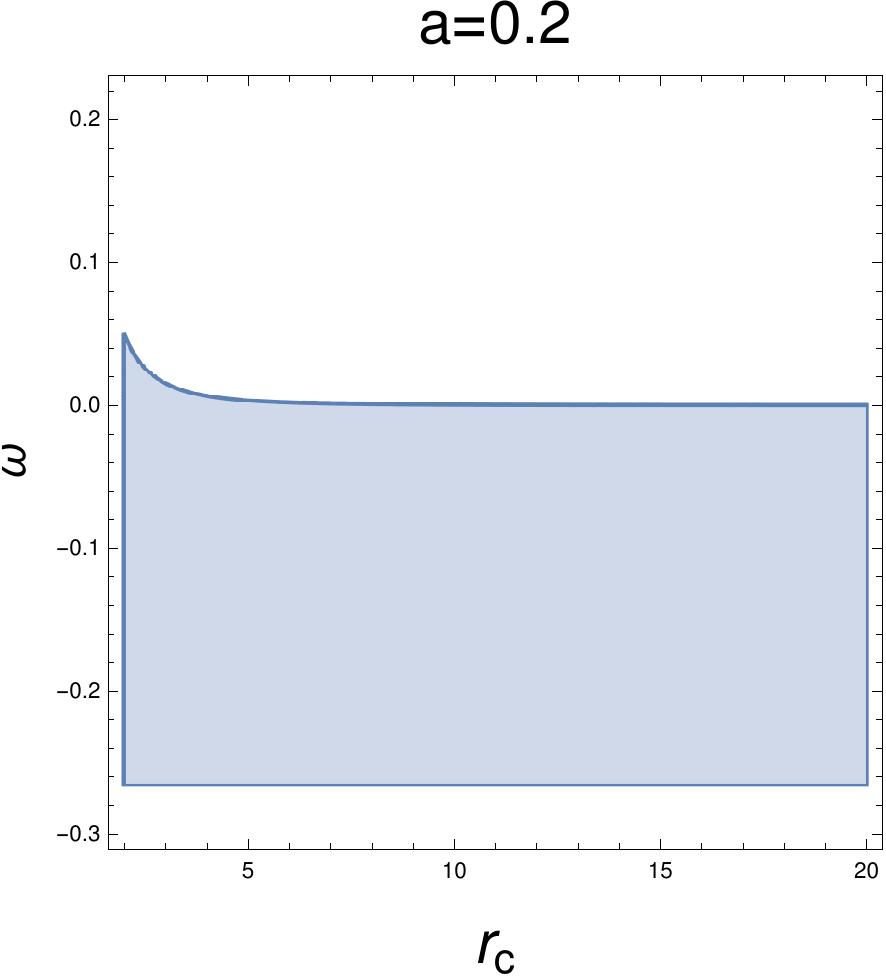} \\
\includegraphics[width=.45\hsize]{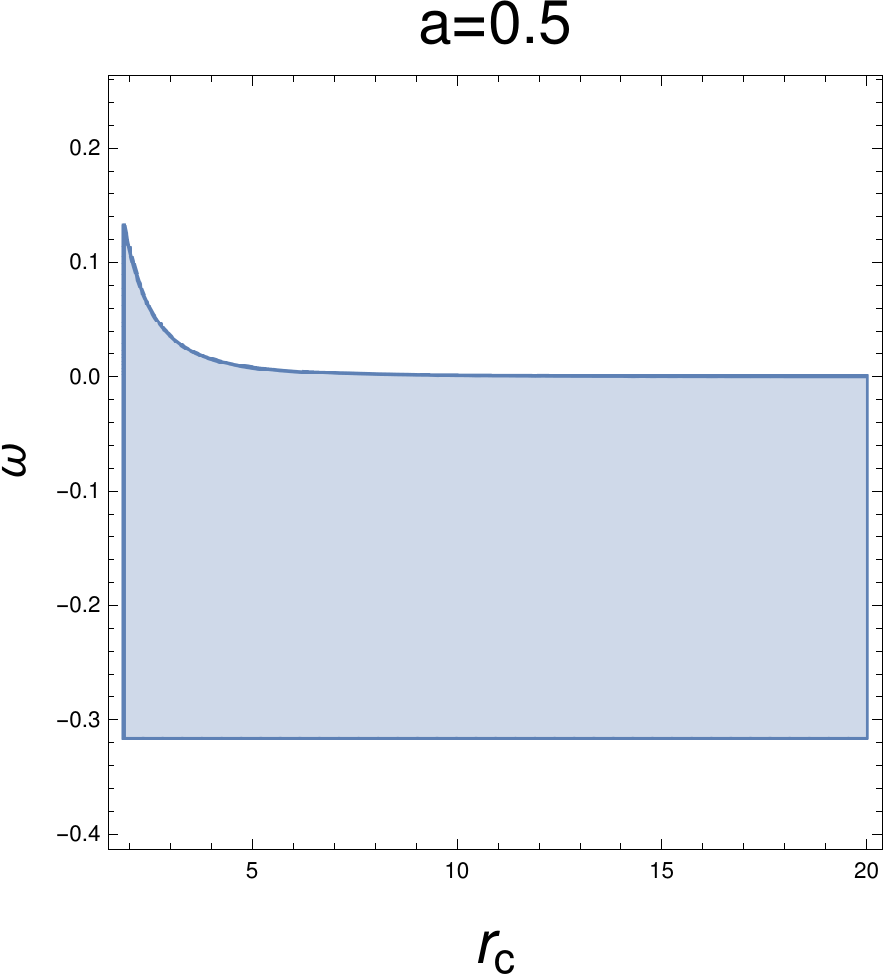} &  \includegraphics[width=.45\hsize]{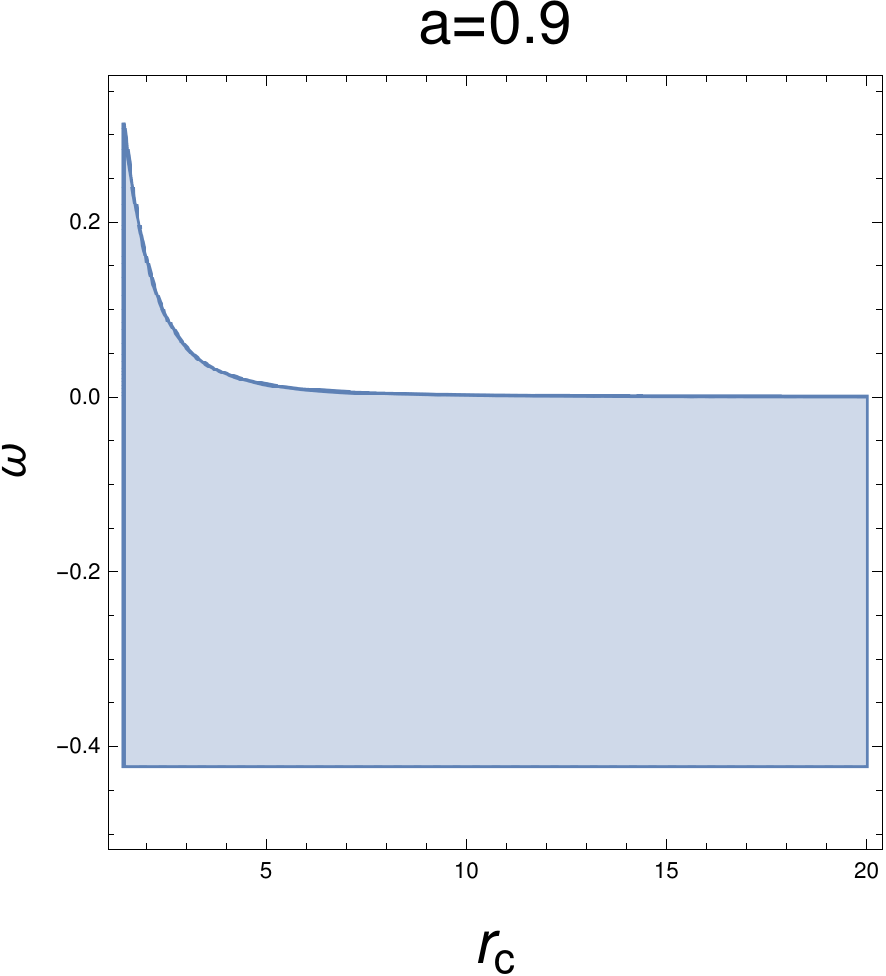} \\
\end{tabular}
	\caption{\label{f4} The shaded region represents $\omega$-values allowed by the sufficient condition 
	\eqref{eq:MaxCond2} for $e=4.17$, $f(S)$ given by Eq. \eqref{eq:fpol} and four values of the spin.}
\end{figure}

Now we search for maxima at the  coordinates ($r_c,\theta_c=0$). We set
\begin{equation}
\label{eq:fpol}
f(S)=-\frac{2k_0}{eB}S,
\end{equation}
where $k_0$ is a constant that has to be chosen, and which represents the effect due to the fluid's charge. The second necessary condition \eqref{eq:ExtremumCond2} is automatically fulfilled for $\theta_c=0$. Using Eqs. \eqref{eq:fpol}, \eqref {eq:drhFin} and the definitions of $U_t$ and $l$, the first necessary condition \eqref{eq:ExtremumCond1} implies
\footnotesize{
\begin{equation}
\label{eq:muPol}
k_0=\frac{1}{2B}\frac{e({a}^2+r_c^2)^2}{(e-2{a})({a}^3+{a}r_c^2+er_c-2{a}r_c)({a}^2+r_c^2-2r_c)}.
\end{equation}}
\normalsize
By choosing the same electromagnetic parameter as in \cite{KovarTr14}, i.e. $e=4.17$ (meaning the same $B$ and $Q$ for comparison with the previous paper), we plot $B k_0$ for various values of $a$ (see Fig.~\ref{f2}); the case of $a=0$ has been fully studied in the article cited above.
Validity of the first sufficient condition \eqref{eq:MaxCond1} is plotted in Fig.~\ref{f3} and summarized in Tab. \ref{tab:cond1}, depending on the value of the spin $a$.
\begin{table}
\caption{\label{tab:cond1} Summary of the validity of the sufficient condition \eqref{eq:ExtremumCond1}, for $e=4.17$}
\begin{ruledtabular}
\begin{tabular}{cc}
$a$   &  $\drr h(r_c,0)<0$ if \\ \hline
0 & $r_c>4$  \\ \hline
0.2 & $r_c>4.47$ \\ \hline
0.5 & $r_c>5.80$ \\ \hline
0.9 & $r_c>16.65$ \\ 
\end{tabular}
\end{ruledtabular}
\end{table}
We note that for $a=0.9$, with the same distribution of charge (i.e. the same $f(S)$ function) and the same value of $e$, equilibrium configurations can exist only relatively far away from the black hole. The second sufficient condition \eqref{eq:MaxCond2} leads to constraints on the angular velocity $\omega$, which is presented in Fig.~\ref{f4}. 
We note that this second sufficient condition gives us another point of view of the $\omega$ value. Note that values of $\omega$ are also restricted by the condition ${\cal{P}}>0$ (see Fig.~\ref{f1}).

\subsection{Equatorial torus case}
\begin{figure}
\begin{tabular}{cc}
\includegraphics[width=.45\hsize]{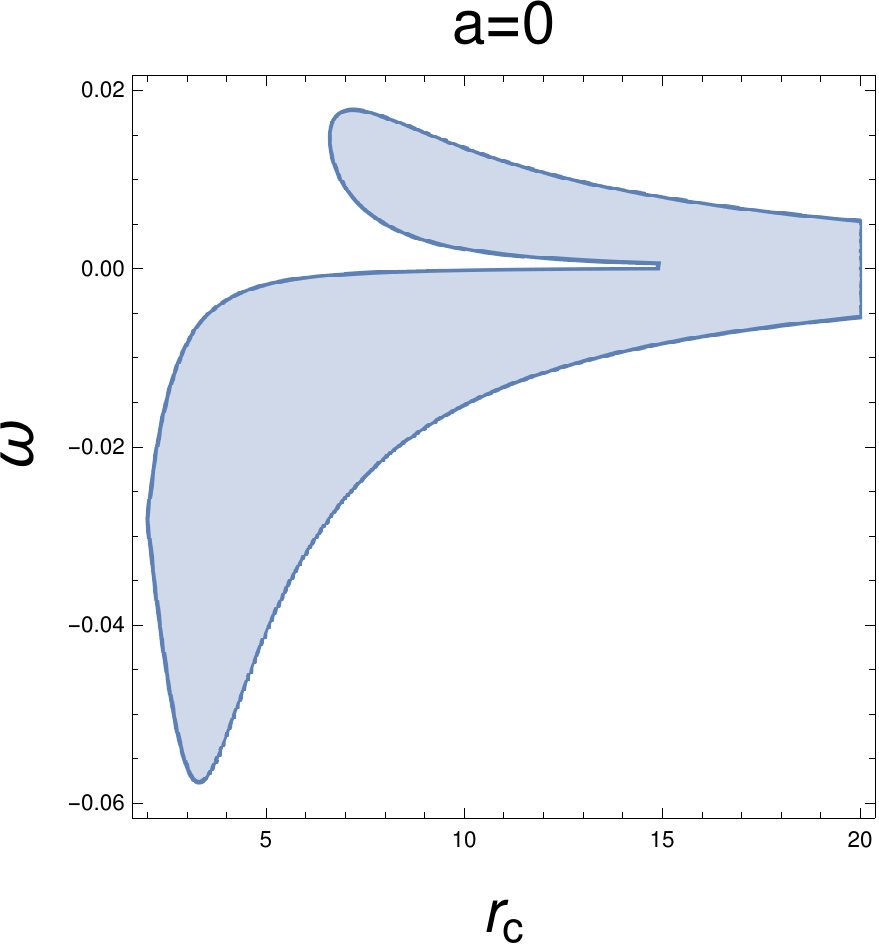} &  \includegraphics[width=.45\hsize]{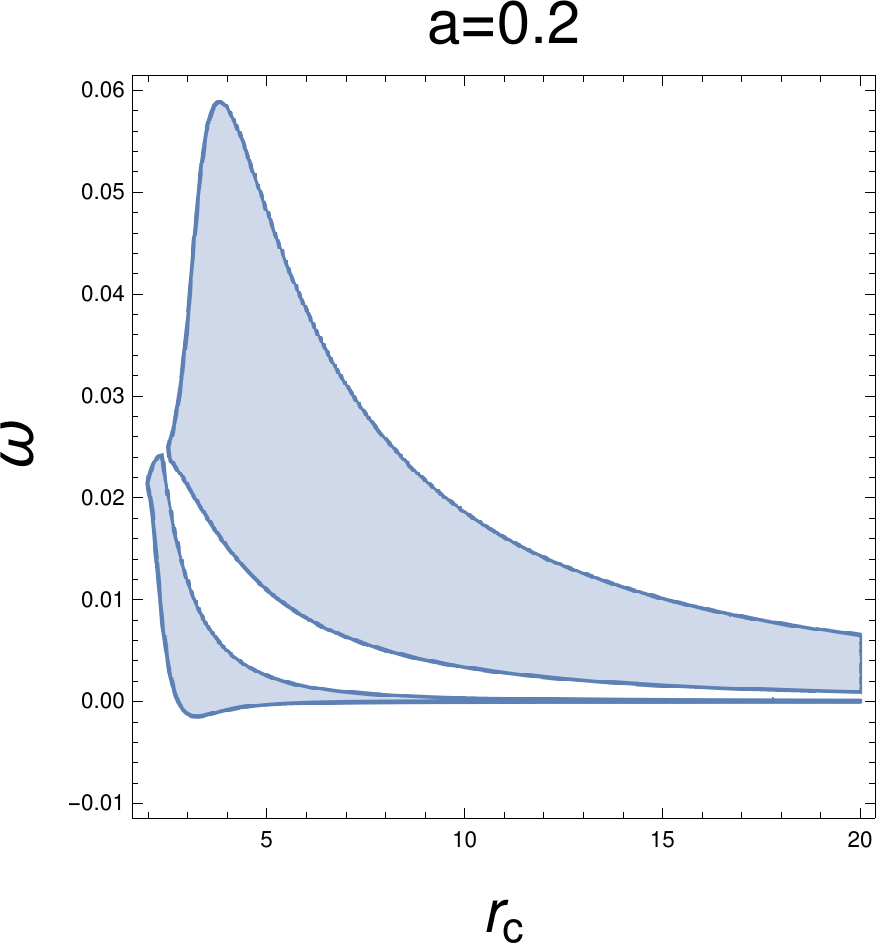} \\
\includegraphics[width=.45\hsize]{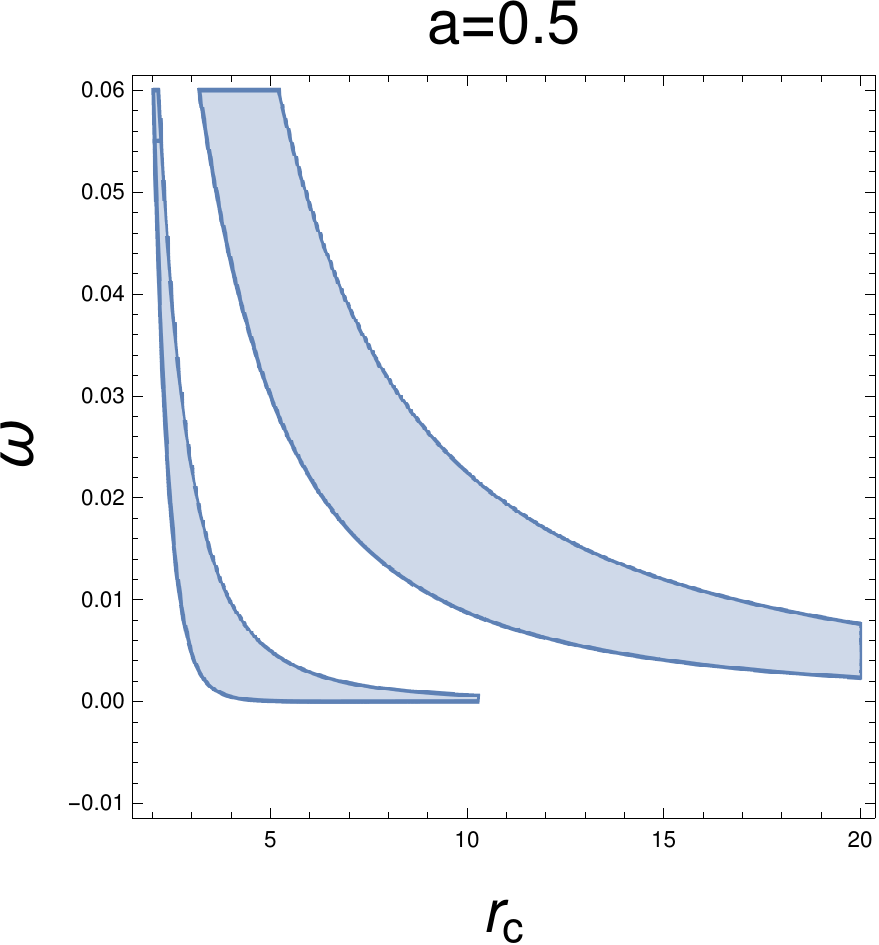} &  \includegraphics[width=.45\hsize]{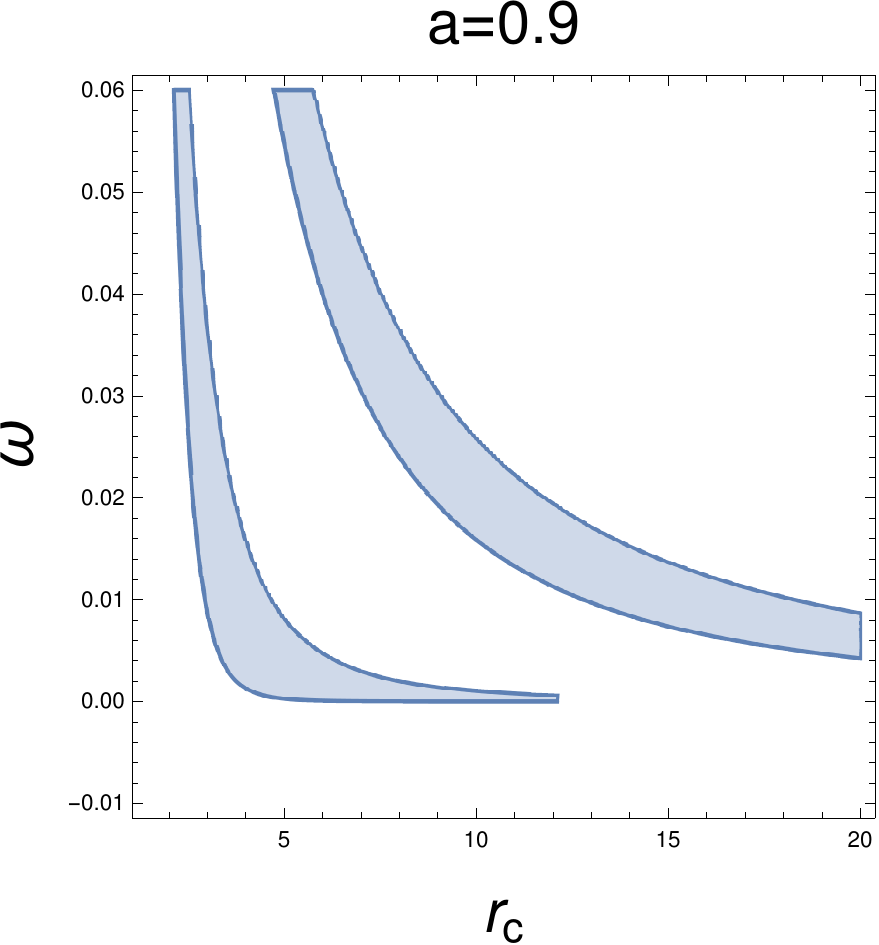} \\
\end{tabular}
 \caption{\label{f5} Union of the conditions \eqref{eq:MaxCond1} and \eqref{eq:MaxCond2} together with the condition $\mathcal{P}>0$. The shaded region represents the area where all conditions are valid, meaning that the value of $\omega$ allows a solution. They are shown for four values of $a=0,0.2,0.5,0.9$.}
\end{figure}

Our second interest is to focus on structures in the equatorial plane. Thus, we search for maxima of $h(r,\theta)$ at the coordinates $(r_c,\theta_c=\pi/2$). Here, we set
\begin{equation}
\label{eq:fEq}
f(S)=\frac{k_0}{4}S^{-2}.
\end{equation} 
The second necessary condition \eqref{eq:ExtremumCond2} is automatically fulfilled for $\theta_c=\pi/2$. The first necessary condition \eqref{eq:ExtremumCond1}, as above, gives us a relation for $k_0=k_0(r_c,a,e,\omega,B)$, which
is now also dependent on $\omega$. The sufficient conditions \eqref{eq:MaxCond1} and \eqref{eq:MaxCond2} for the existence of equatorial tori and the basic necessary condition $\mathcal{P}>0$ are displayed in Fig.~\ref{f5}. We note that for large spin ($a=0.5$ and $a=0.9$), only co-rotating tori ($\omega>0$) are possible. For small $r_c$, we can see that there is no solution within the ergosphere.

We note that there is a clear difference between the conditions of existence for the case of non-rotating black holes and the rotating ones. \\

Of course, both the represented cases are dependent on the function $f(S)$ and the value of $e$. If we change one of these parameters, the conditions of existence will be different.

\section{Construction of equilibrium configurations \label{sec:section5}}
Once the conditions of existence are known, meaning that we can choose the right values of $\omega$ and $k_0$ satisfying conditions \eqref{eq:ExtremumCond} -- \eqref{eq:MaxCond}, we can construct the equilibrium configurations. As said in \cite{KovarTr14}, the configurations highly depend on the values of $B$ and $Q$. The electromagnetic field has to be strong enough to compete with the gravitational field of the black hole. Secondly, to fulfill the assumed test field approximation, $Q$ and $B$ have to be higher than the total charge and the magnetic field  produced by the torus itself. As our uniform magnetic test field is assumed to be produced by a distant magnetic source with a dipole field, the maximum strength is $\tilde{B} \simeq 10^6$ T, corresponding to $B \simeq 10^{-11}$. To check our weak self-field approximation, we can calculate the total charge of the torus, determined by the following equation:
\begin{equation}
{\cal{Q}}=\int_V qdV,
\end{equation}
and the magnetic field generated by the torus itself. We approximate the torus by an infinitely thin ring centered in $r=r_c$. The strength of the self-${\cal{B}}$ field close to the outer edge of the torus can be approximated by \cite{KovarTr14}
\begin{equation}
\tilde{{\cal{B}}}=\frac{\tilde{\mu_0}\tilde{{\cal{Q}}}\tilde{\omega}}{4\pi(\tilde{r}_{\text{out}}-\tilde{r}_c)},
\end{equation}
where $r_{\text{out}}$ is the outer edge radius and $\mu_0$ the vacuum permeability.

\subsection{Polar cloud configuration}
\begin{figure*}
\centering
\begin{tabular}{cc}
\includegraphics[width=.45\hsize]{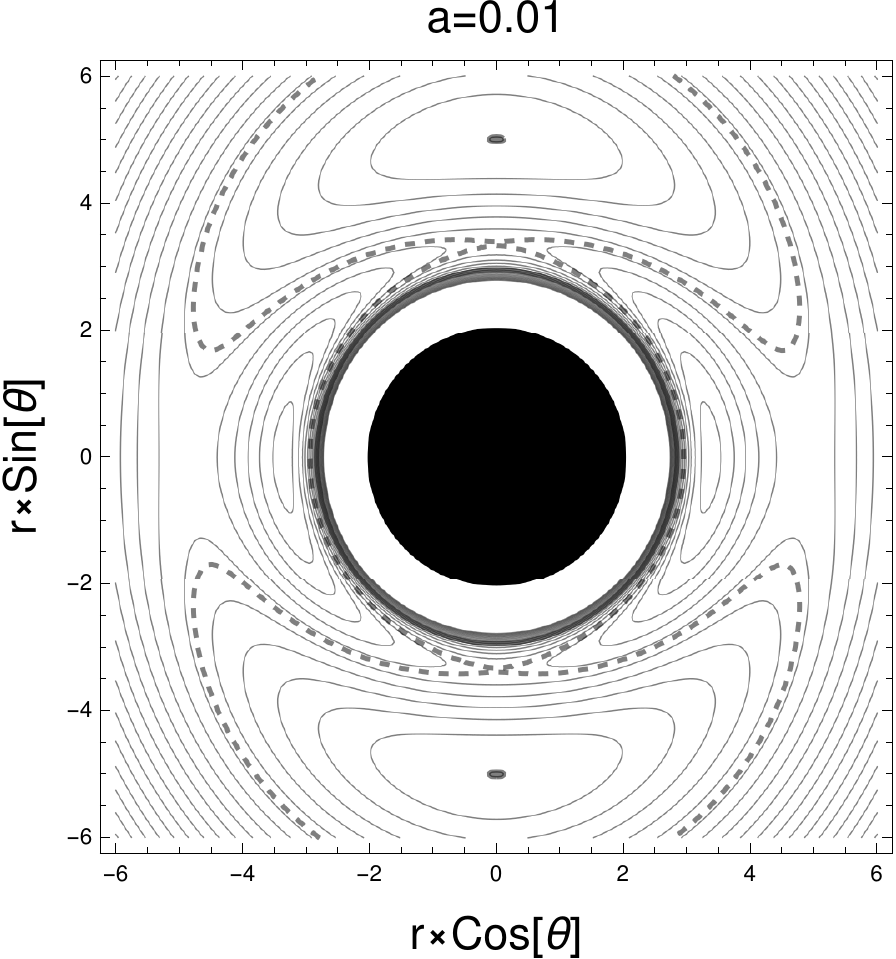} &  \includegraphics[width=.45\hsize]{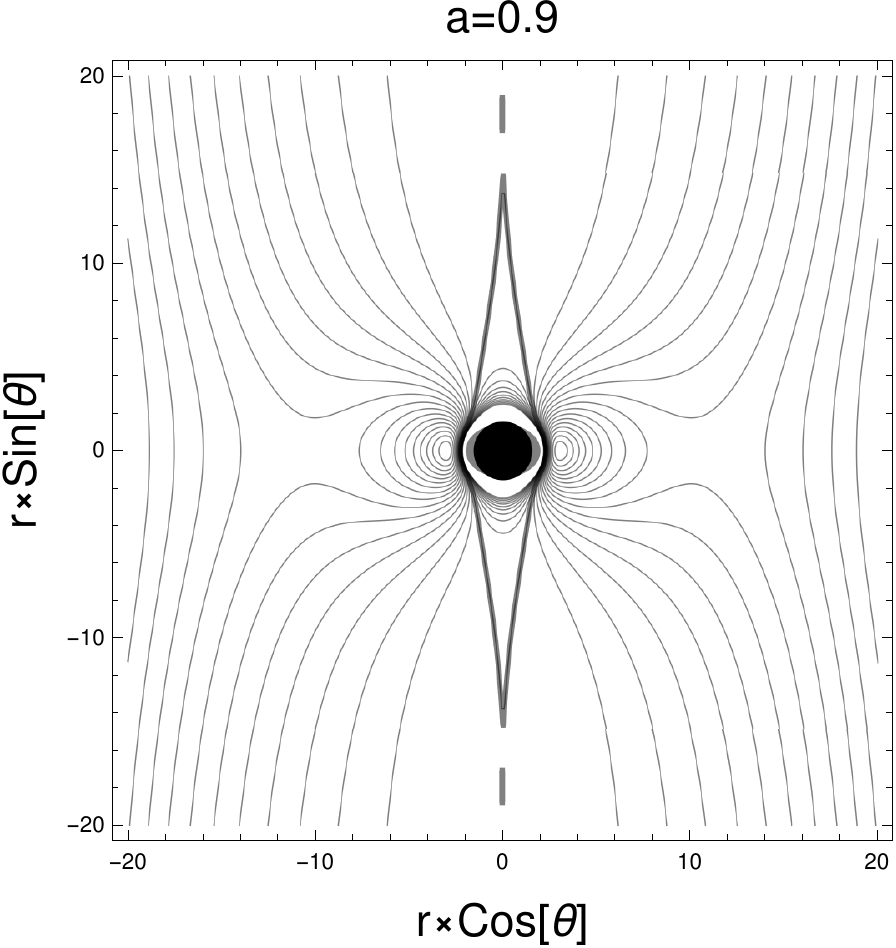}
\end{tabular}
\caption{\label{f7} Poloidal sections of equipressure surfaces for a slowly rotating black hole on the left ($a=0.01$) and a fast rotating black hole on the right ($a=0.9$). The thick curves mark, where the $h$-function becomes zero (i.e zero pressure and zero density). The tiny clouds are centered on the z-axis at $(r=5,\theta=0)$ (on the left) and at $(r=18,\theta=0)$ (on the right). The dashed line represents the equipotential leading to a cusp point (cannot be resolved on the right graphic due to the chosen scale). The gray region represents the ergosphere. In the plot at the left, the ergosphere cannot be seen, since it is nearly completely swallowed by the black hole horizon.}
\end{figure*}

\begin{figure*}
\begin{tabular}{cc}
\includegraphics[width=.35\hsize]{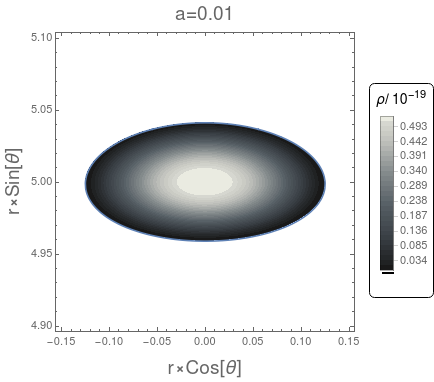} &  \includegraphics[width=.35\hsize]{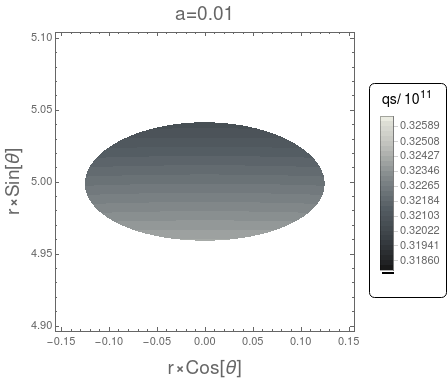} \\
\includegraphics[width=.35\hsize]{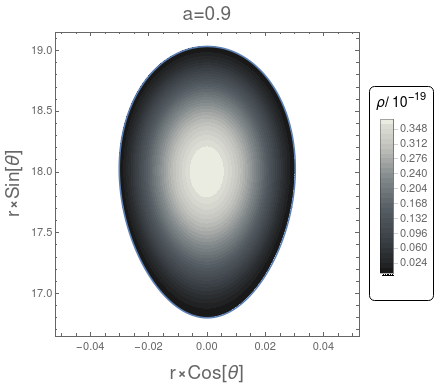} &  \includegraphics[width=.35\hsize]{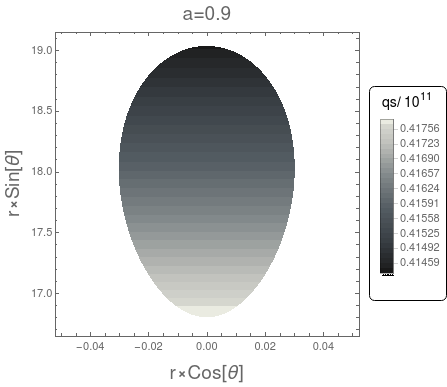} 
\end{tabular}
\caption{\label{f8} Mass (on the left left) and specific (on the right right) charge density profile of the polar clouds. The top panels correspond to the case of slow rotation ($a=0.01$) from the configuration shown on the left side of Fig.\ref{f7}. The bottom panels are for the case of fast rotation ($a=0.9$), as shown on the right side of Fig.\ref{f7}.}
\end{figure*}

In this section, we are going through the construction of the polar cloud configurations. We show solutions for two different values of the black hole spin, $a=0.01$ (slow rotation) and $a=0.9$ (fast rotating black hole). In the first case, we set the following parameters, $B=10^{-11}$ (corresponding to $\tilde{B}\doteq 2.3\times 10^{6}$ T), $e=Q/B=4.17$ (corresponding to $Q=4.17 \times 10^{-11}$ and $\tilde{Q}\doteq 7.17\times 10^{9}$ C), $f(S)$ given by Eq. \eqref{eq:fpol}, $r_c=5$ and $\omega=-1/1000$. According to these values and Eq. \eqref{eq:muPol}, $k_0$ has to be set to $k_0B\doteq 0.9976$ in order to find a solution. The integration constant is set to $h_0\doteq -0.03549$, which defines the boundary of the torus (see Fig.~\ref{f7}).
By using the $h$-function, through Eqs. \eqref{eq:hLinkp}, \eqref{eq:pressure}, \eqref{eq:density} and \eqref{eq:specificCharged}, we can determine the physical characteristics of the torus, such as the pressure, the mass density and the charge density. In our case, we set $\Gamma=5/3$ and $\kappa=10^7$, which leads to the mass density and the specific charge distributions shown in Fig.~\ref{f8} (top). 
\begin{table}
\caption{\label{tab:PhyValPolSlow} Maximal values of the physical characteristics corresponding to the polar cloud described in Fig.~\ref{f8} (top).}
\begin{ruledtabular}
\begin{tabular}{ll}
$\rho_{\text{max}}$    $\simeq 10^{-20}$  &  $\tilde{\rho}_{\text{max}}$  $\simeq 10$ kg$\cdot$ m$^{-3}$\\ 
$p_{\text{max}}$    $\simeq  10^{-25}$ &  $\tilde{p}_{\text{max}}$  $\simeq  10^{12}$ Pa  \\ 
$q_{\text{max}}$    $\simeq  10^{10}$  &  \\
\end{tabular}
\end{ruledtabular}
\end{table}
Their maximal values are summarized in Tab. \ref{tab:PhyValPolSlow}. We can also compute the total charge of the torus, ${\cal{Q}} \simeq 10^{-13}$ (corresponding to $\tilde{{\cal{Q}}}\simeq 10^8$ C), which is almost two orders lower than the charge of the black hole. This gives us the strength of the magnetic field ${\cal{B}} \simeq 10^{-16}$ (corresponding to $\tilde{{\cal{B}}} \simeq 1$ T) close to the outer edge. Thus, we can conclude that our weak self-field approximation is valid in this case.

As we can see in Fig.~\ref{f8} the morphology of the solution is basically the same as in the non-rotating case, with an ellipsoidal shape and a tiny cross-section ($\rin\doteq4.959, \rout\doteq5.041$).

We turn now to the second case of a fast, rotating black hole ($a=0.9$). In this configuration, according to Fig.~\ref{f3}, the center of the torus has to be located farther away from the black hole than in the previous case. We set $r_c=18$, and take the same values as before for $B$, $e$ and $\omega$, which leads to a new value of $k_0 B\doteq 0.9601$. Furthermore, $h_0$ is set to $h_0 \doteq 0.0745$. The chosen parameters involve the configuration presented in Fig.~\ref{f7} (right). As before, setting $\Gamma=5/3$ and $\kappa=10^6$, we obtain the mass and specific charge density profiles described in Fig.~\ref{f8} (bottom). 
We can see, the resulting configuration is different as compared to the first case, as it is now close to an oblate ellipsoid. The maximal values of its physical characteristics are given in Tab. \ref{tab:PhyValPolFast}.
\begin{table}
\begin{center}
\caption{\label{tab:PhyValPolFast}  Maximal values of the physical characteristics of the torus corresponding to the polar cloud described in Fig.~\ref{f8} (bottom].}
\begin{ruledtabular}
\begin{tabular}{ll}
$\rho_{\text{max}}$   $\simeq 10^{-20}$  &  $\tilde{\rho}_{\text{max}}$ $\simeq 10$ kg$\cdot$ m$^{-3}$\\ 
$p_{\text{max}}$  $\simeq 10^{-27}$ &  $\tilde{p}_{\text{max}}$ $\simeq 10^{11}$ Pa  \\ 
$q_{\text{max}}$   $\simeq 10^{10}$  &   \\
\end{tabular}
\end{ruledtabular}
\end{center}
\end{table}
Again, we have to check the validity of the weak self-field approximation. The total charge of the rotating polar cloud, ${\cal{Q}} \simeq 10^{-12}$ (corresponding to $\tilde{{\cal{Q}}}\simeq10^8$ C), is again almost two orders lower than the charge of the black hole. The associated magnetic field ${\cal{B}} \simeq 10^{-17}$ (corresponding to $\tilde{{\cal{B}}}\simeq0.12$ T) is sufficiently weak to validate our assumption.
 
\subsection{Equatorial tori configuration}
\begin{figure*}
\centering
\begin{tabular}{cc}
\includegraphics[width=.45\hsize]{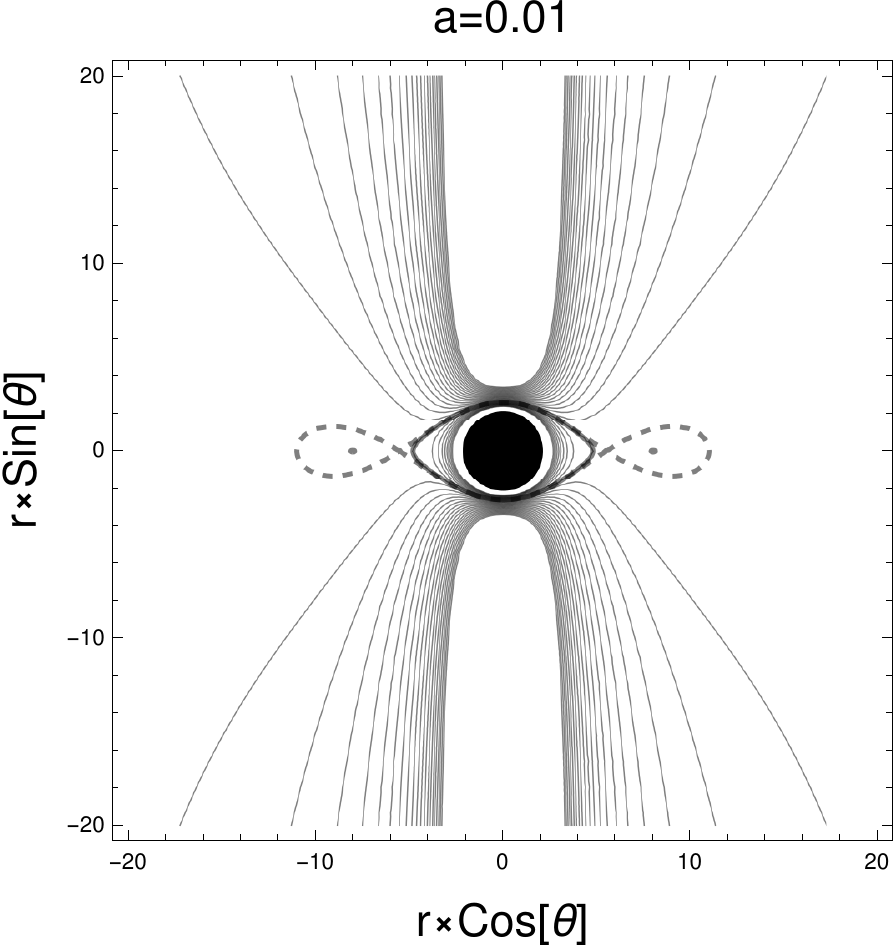} &  \includegraphics[width=.45\hsize]{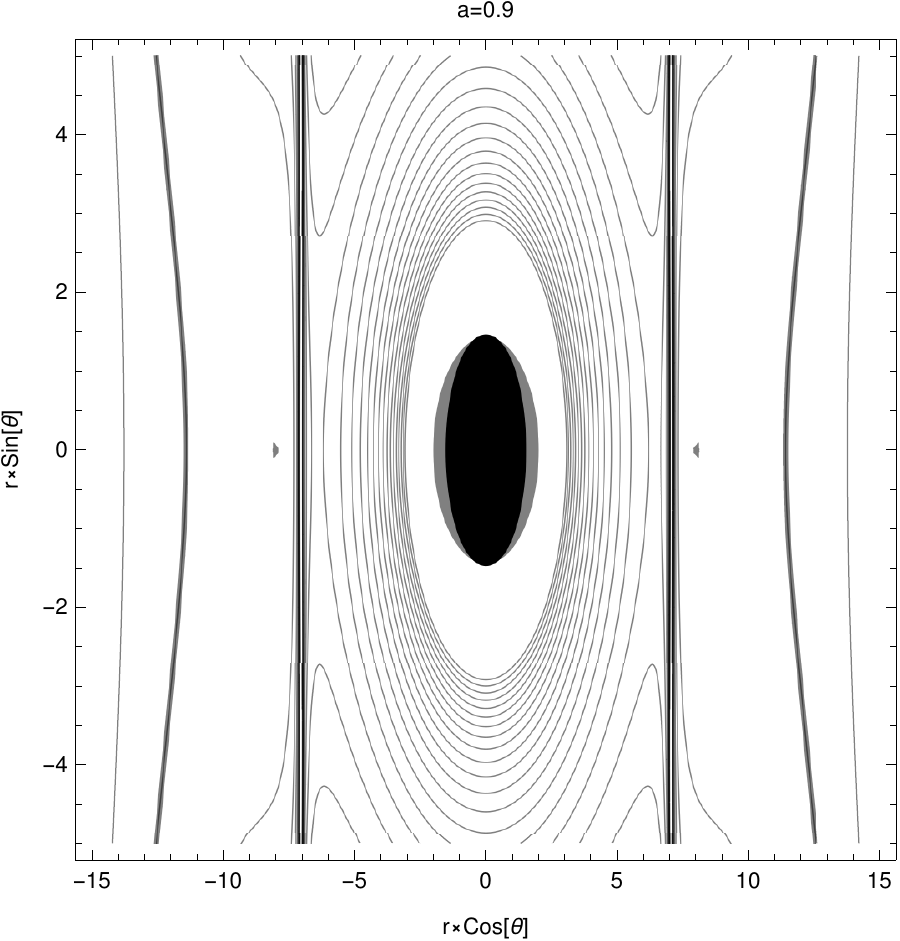}
\end{tabular}
\caption{\label{f9} Equatorial tori configurations for a slowly rotating black hole at the left ($a=0.01$) and a fast rotating black hole ($a=0.9$). The thick curves mark, where the $h$-function becomes zero (i.e zero pressure and zero density). The tiny tori are located at $(r=8,\theta=\pi /2$) for both configurations. The dashed line represents the equipotential leading to a cusp point (cannot be resolved on the right graphic due to the chosen scale) The gray region represents the ergosphere. In the plot at the left, the ergosphere cannot be seen, since it is nearly completely swallowed by the black hole horizon.}
\end{figure*}

\begin{figure*}
\begin{tabular}{cc}
\includegraphics[width=.35\hsize]{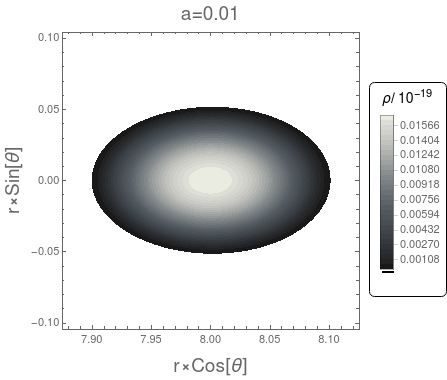} &  \includegraphics[width=.35\hsize]{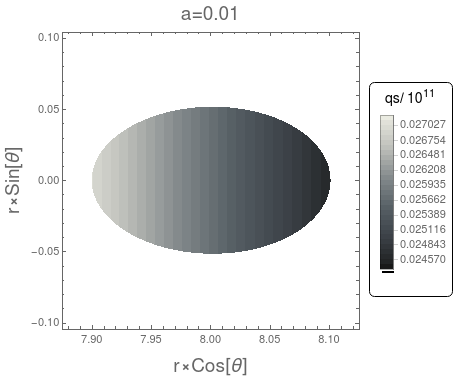} \\
\includegraphics[width=.35\hsize]{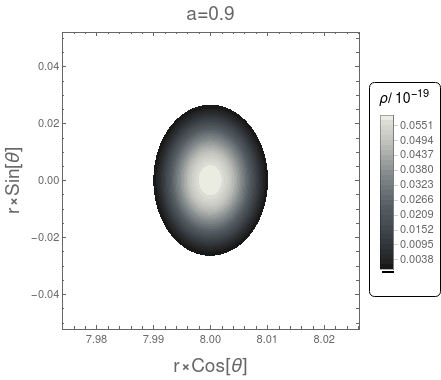} &  \includegraphics[width=.35\hsize]{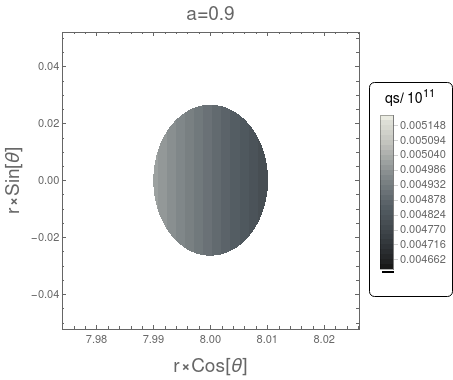} 
\end{tabular}
\caption{\label{f10}Mass (on the left left) and specific (on the right right) charge density profile of the polar clouds. The top panels correspond to the case of slow rotation ($a=0.01$) from the configuration shown on the left side of Fig.\ref{f10}. The bottom panels are for the case of fast rotation ($a=0.9$), as shown on the right side of Fig.\ref{f10}.}
\end{figure*}

The second interesting case, we will have a look at, is about tori configurations in the equatorial plane. We perform a similar study to the one made for the polar clouds, and analyze solutions for the case of a slowly ($a=0.01$) and a fast rotating black hole ($a=0.9$). In the case of a slowly rotating black hole, we set the parameters of choice as follows: $r_c=8$, $B=8.78 \times 10^{-11}$ (corresponding to $\tilde{B}=2.35 \times 10^4$ T), $e=-1/8.78$ (corresponding to ${Q}=-10^{-11}$ and $\tilde{Q}=-1.72 \times 10^9$ C) and $\omega=1/100$. According to these parameters and the existence conditions, we get $k_0 \doteq 0.10819B$. Finally, we set $h_0\doteq-0.025995$. The corresponding equatorial torus is represented in Fig.~\ref{f9}.
As before, to obtain the mass density and the specific charge profile, we set the polytropic index to $\Gamma=5/3$ and the polytropic coefficient to $\kappa=10^8$. This is shown in Fig.~\ref{f10}. The physical characteristics are given in Tab. \ref{tab:PhyValEqSlow}.
\begin{table}
\begin{center}
\caption{\label{tab:PhyValEqSlow}  Maximal values of the physical characteristics of the torus corresponding to the polar cloud described at the top of Fig.~\ref{f9}.}
\begin{ruledtabular}
\begin{tabular}{ll}
$\rho_{\text{max}}$   $\simeq  10^{-21}$  &  $\tilde{\rho}_{\text{max}}$ $\simeq 1$ kg$\cdot$ m$^{-3}$\\ 
$p_{\text{max}}$   $\simeq 10^{-27}$ &  $\tilde{p}_{\text{max}}$  $\simeq 10^{10}$ Pa  \\ 
$q_{\text{max}}$   $\simeq  10^{9}$  & \\
\end{tabular}
\end{ruledtabular}
\end{center}
\end{table}
We note that in the case of a slow rotation of the central mass, the morphology of the solution does not change. We obtain a rotating torus with a tiny equatorial cross-section ($\rin\doteq7.9174,\rout\doteq8.0836$). In the final step, we check the weak self-field approximation again: the total charge of the torus, ${\cal{Q}}\simeq 10^{-13}$ (corresponding to $\tilde{\cal{Q}}\simeq 10^7$ C) is two orders lower than the charge of the black hole, and the magnetic field generated by the torus is approximately to ${\cal{B}}\simeq10^{-14}$ (corresponding to $\tilde{{\cal{B}}}\simeq 10^2$ T). 

In case of a fast rotating central mass, according to Fig.~\ref{f5}, we choose the center of the torus to be at $r_c=8$ again. If we take the same values as above for $B$ and $e$, according to Fig.~\ref{f5}, we can choose $\omega=0.03$, which leads to $k_0\doteq 0.007329B$ for the given set of parameters. The equatorial torus corresponding to these parameters is shown in Fig.~\ref{f9} (right). Using $\Gamma=5/3$ and $\kappa=10^7$, the associated mass and the specific charge density are shown in Fig.~\ref{f10} (bottom). The maximal values are given in Tab. \ref{tab:PhyValEqFast}. 
\begin{table}
\begin{center}
\caption{\label{tab:PhyValEqFast}  Maximal values of the physical characteristics of the torus corresponding to the polar cloud described at the bottom of Fig.~\ref{f9}.}
\begin{ruledtabular}
\begin{tabular}{ll}
$\rho_{\text{max}}$   $\simeq  10^{-21}$  &  $\tilde{\rho}_{\text{max}}$ $\simeq 2$ kg$\cdot$ m$^{-3}$\\ 
$p_{\text{max}}$   $\simeq 10^{-27}$ &  $\tilde{p}_{\text{max}}$ $\simeq 10^{10}$ Pa  \\ 
$q_{\text{max}}$   $\simeq 10^{8}$  &  $\tilde{q}_{\text{max}} \simeq$  \\
\end{tabular}
\end{ruledtabular}
\end{center}
\end{table}
As in the previous polar cloud configuration, the torus has an oblate spheroidal shape. The total charge of the torus, ${\cal{Q}}\simeq 10^{-14}$ (corresponding to $\tilde{\cal{Q}}\simeq 10^6$ C) is three orders lower than the charge of the black hole, and the magnetic field generated by the torus is approximately to ${\cal{B}}\simeq 10^{-13}$ (corresponding to $\tilde{{\cal{B}}}\simeq10^2$ T). 

\section{Limiting cases}
After a general study, we focus our attention on two limiting cases: (i) $Q=0, B \neq 0$ and (ii) $B=0, Q \neq 0$. In this part our interest is only on the condition of existence of polar clouds. In \cite{KovarTr14}, with a non-rotating central object and a charged perfect fluid in permanent rigid rotation, no polar cloud was found. By contrast, this is now possible for a certain choice of $f(S)$ and certain sets of parameters. 
\subsubsection{$Q=0, B \neq0$}
After some tests of $f$-functions, we set $f(S)=-k_0/S^2$. The first necessary condition of existence \eqref{eq:ExtremumCond1} implies the following condition on $k_0(r_c,a,B)$,
\begin{equation}
k_0=-\frac{1}{2}B\frac{{a}\left(r_c^2+{a}^2-2r_c\right)}{r_c^2+{a}^2}.
\end{equation}
As in the general case in section \ref{sec:ConditionPolarCloud}, the second necessary condition of existence \eqref{eq:ExtremumCond2} is automatically fulfilled. The second sufficient condition leads to constraints on the choice of $\omega$ which has to be as 
$\omega \in [1/a;2ar_{\rm c}/\left(r_{\rm c}^2+{a}^2\right)^2]$. Finally, the first sufficient condition \eqref{eq:MaxCond1} implies 
\begin{equation}
-\frac{2\left(r_c^2+{a}^2\right)^2}{\left(r_c^2+{a}^2-2r_c\right)\left(r_c^2+{a}^2\right)}<0,
\end{equation}
which is always true if $r_c^2+{a}^2-2r_c>0$, i.e. if $r_c>r_h$, which is always the case. This condition is satisfied by the conditions ${\cal{P}}>0$ analysed in Sec. \ref{sec:CondPPos}. Thus, polar clouds can exist with an uncharged rotating compact object. 

\subsubsection{$B=0, Q\neq 0$}
Setting $f(S)=-2 k_0 S$, Eq. \eqref{eq:ExtremumCond1} gives us $k_0(r_c,a,Q)$,
\begin{equation}
k_0=-\frac{1}{2Q^2}\frac{\left(r_c^2+{a}^2\right)^2}{\left(r_c^2+{a}^2-2r_c\right)r_c}.
\end{equation}
The second necessary condition of existence \eqref{eq:ExtremumCond2} is automatically satisfied. The second sufficient condition \eqref{eq:MaxCond2} implies $\omega \in [0;2ar_c/\left(r_c^2+{a}^2\right)^2]$. Finally, the first sufficient condition \eqref{eq:MaxCond1} is represented in Fig.~\ref{f11} for various values of the spin parameter.
\begin{figure}
\centering
  \includegraphics[width=0.7\hsize]{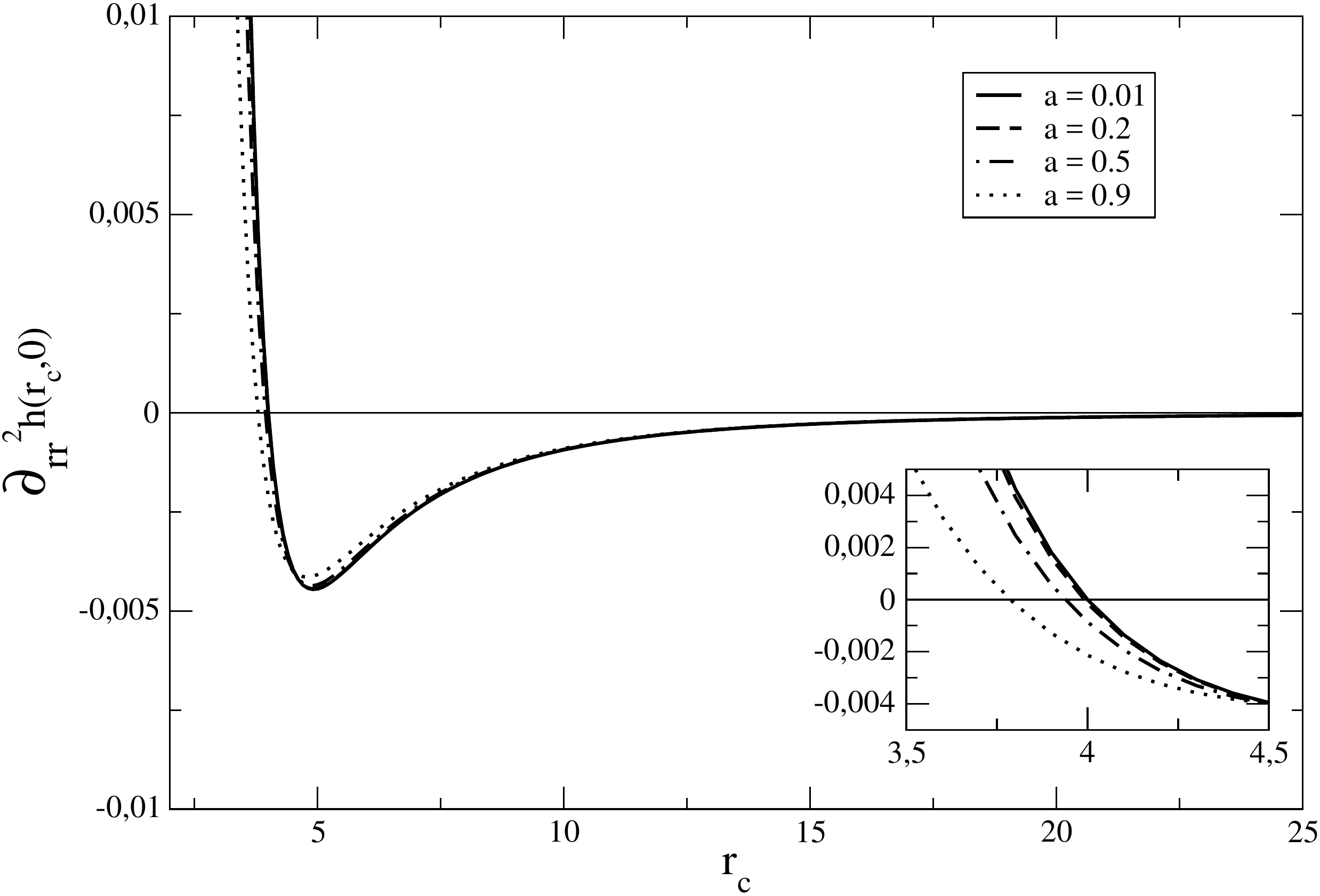}
	\caption{\label{f11} Behavior of $\drr h(r_c,0)$ for four values of the spin $a=0.01,0.2,0.5,0.9$, for $Q\neq 0,B=0$ and $f(S)=-2 k_0 S$.}
	\label{f11}
\end{figure}
We conclude that also in this second limit, equilibrium polar clouds can be found. Thus, it seems that the rotation of the compact object plays an important role in the existence of such structures.

\section{Conclusion}
In this paper, we presented the model of equilibrium configurations of electrically charged perfect fluids encircling a rotating (Kerr) black hole endowed with test charge and embedded into a large scale asymptotically uniform magnetic field. The introduced work can be compared to the preceding one \cite{KovarTr14}, where the central black hole is considered non-rotating and, as well as here, the structures like polar clouds and equatorial tori are discussed. 

We started our study by analyzing the conditions of existence of both the types of configurations for various values of the black hole spin. As the study depends on many parameters, such as the black hole charge, the strength of the uniform magnetic field, the black hole spin, the angular velocity and the distribution of charge throughout the torus, we decided to fix the latter one and used the conditions of existence to constrain the others. The most important conclusion, that can be drawn, is that for both the configurations, the polar clouds and the equatorial tori, the addition of a non-zero spin seriously changes the existence conditions and the morphology of the solution. In the case of polar clouds, and for a fast rotating black hole (high spin), solutions can only be found relatively far away from the black hole, which is not the case for slow rotation (small spin). The same result was found for the equatorial tori, where we can see that the range of possible solutions substantially changes with the spin value. We have to note that the conditions plotted in Sec. \ref{sec:section4} are valid only for the distribution of charge we chose. They will be different if another distribution is used.

In the second part of our study, Sec. \ref{sec:section5}, we focused on the construction of polar clouds and equatorial tori. We discussed in details two different cases: the one for a slowly rotating black hole and the one for a fast rotating black hole. Each of these cases was tested for both the considered types of configurations. We showed that the density, the pressure profile and the morphology highly depend on the strength of the magnetic field but also on the spin of the black hole. We also found that for both the types of configurations, for a selected set of parameters, in the case of fast rotation, the morphology of the structures is close to an oblate shape.

Finally, by analyzing two limiting cases, i.e. the zero test electric charge of the black hole and the zero magnetic field strength, respectively, we can say that the rotation plays very important factor on the possibility of the presence of the equilibrium polar configurations. 
In the case of a non-rotating black hole, the existence of polar clouds needs both the external magnetic field and the test electric charge of the black hole. The situation changes for a rotating central black hole, which allows for the existence of polar clouds also when one of the background parameters, i.e. the black hole charge or magnetic field strength, is set to zero. 

Considering following studies of the charged fluid equilibrium configurations, along with the more detailed discussion of the possible morphologies in dependence on the chosen charge density distributions, we can see a challenge for an addition of the self-fields (gravitational and magnetic) into our model. In this respect, for now, the constructed configurations must have been kept with very small cross-section. The consideration of the self-fields could allow us to obtain more extended solutions imitating real accretion discs.

\begin{acknowledgements}
The authors VK, JK and PS acknowledge the project ``Albert Einstein Center for Gravitation and Astrophysics" - Czech Science Foundation GA\v{C}R No. 14-37086G. JK and PS would also like to express their acknowledgment for the Institutional support of the Faculty of Philosophy and Science, Silesian University in Opava. The authors AT, KS and EH acknowledge support from the Research Training Group 1620 ``Models of Gravity" funded by the German Science Foundation DFG. KS and EH are also grateful for support from the DFG funded Collaborative Research Center 1128 `Relativistic Geodesy and Gravimetry with Quantum Sensors (geo-Q)'.
\end{acknowledgements}
\bibliographystyle{apsrev4-1}
\bibliography{mybibtex}

\end{document}